\documentclass[structabstract]{aa}
\usepackage{savesym}
\usepackage{amsmath}
\savesymbol{bibfont}
\usepackage{graphicx}
\usepackage{natbib}
\usepackage{color}
\usepackage{txfonts}
\def\be{\begin{equation}}
\def\ee{\end{equation}}
\def\bdm{\begin{displaymath}}
\def\edm{\end{displaymath}}

\begin{document}
   \title{Dual Maxwellian-Kappa modelling of the solar wind electrons:\\
	new clues on the temperature of Kappa populations}
		\titlerunning{Maxwellian-Kappa solar wind}
   \author{M.\ Lazar \inst{1,2} \fnmsep\thanks{\email{mlazar@tp4.rub.de}} V.\ Pierrard \inst{3,4} 
	S.M.\ Shaaban \inst{1,5} H.\ Fichtner \inst{2,6} \and S.\ Poedts \inst{1} 
    }
         \authorrunning{M.\ Lazar et al.}
   \institute{
	$^1$ Centre for Mathematical Plasma Astrophysics, Celestijnenlaan 200B, 3001 Leuven, 
	Belgium\\	
	$^2$ Institut f\"ur Theoretische Physik, Lehrstuhl IV: Weltraum- und Astrophysik, 
	Ruhr-Universit\"at Bochum, D-44780 Bochum, Germany\\
  $^3$ Royal Belgian Institute for Space Aeronomy, 3 av. Circulaire, B-1180 Brussels, Belgium \\
	$^4$ Universit\'e Catholique de Louvain, Georges Lema\^ itre Centre for Earth and Climate 
	Research (TECLIM), Place Louis Pasteur 3, 1348 Louvain-La-Neuve, Belgium\\
	$^5$ Theoretical Physics Research Group, Physics Department, Faculty of Science, Mansoura University, 35516, Egypt\\ 
	$^6$ Research Department of Complex Plamas, Ruhr-Universit\"at Bochum, D-44780 Bochum, Germany
                }
   \date{Received , 2016; accepted , }

   \abstract
   {Recent studies on Kappa distribution functions invoked in space plasma 
	applications have emphasized two alternative approaches which may assume the 
	temperature parameter either dependent or independent of the power-index 
	$\kappa$. Each of them can obtain justification in different scenarios involving
	Kappa-distributed plasmas, but direct evidences supporting any of these two
	alternatives with measurements from laboratory or natural plasmas are not 
	available yet.}
   {This paper aims to provide more facts on this intriguing issue from direct 
	fitting measurements of suprathermal electron populations present in the solar wind, 
	as well as from their destabilizing effects predicted by these two alternating approaches.}
   {Two fitting models are contrasted, namely, the global Kappa and the dual Maxwellian-Kappa models, 
	which are currently invoked in theory and observations. The destabilizing 
	effects of suprathermal electrons are characterized on the basis of a kinetic approach which accounts 
	for the microscopic details of the velocity distribution.}
   {In order to be relevant, the model is chosen to accurately reproduce the observed distributions 
	and this is achieved by a dual Maxwellian-Kappa distribution function.
	A statistical survey indicates a $\kappa$-dependent temperature of the suprathermal (halo) 
	electrons for any heliocentric distance. Only for this approach the instabilities driven by the temperature 
	anisotropy are found to be systematically stimulated by the abundance of suprathermal populations, i.e., 
	lowering the values of $\kappa$-index. }
   {}
	

   \maketitle
%

\section{Introduction}

Kappa modelling is a widely exploited technique for space plasma diagnosis and kinetic 
analysis \citep{Pierrard-Lazar-2010, Livadiotis-McComas-2013}. The Kappa ($\kappa$-)distribution 
function is nearly Maxwellian at low energies and decreases as a power-law at higher 
energies, enabling a satisfactory overall (global) description of the velocity 
distributions of plasma particles measured in the solar wind \citep{Vasyliunas-1968,
Christon-etal-1989, Collier-etal-1996, Maksimovic-etal-1997}. The power-index 
$\kappa$ quantifies the presence of suprathermal (non-Maxwellian) 
populations, which enhance the high energy tails of the distributions. 
Alternatively, the same Kappa power-laws have also been used to reproduce only 
the high-energy tails (also known as the halo component), while the core
of the distribution is well fitted by a standard Maxwellian \citep{Maksimovic-etal-2005, 
Stverak-etal-2008, Pierrard-etal-2016}.

Recent studies on global Kappa models have emphasized two distinct alternatives, 
namely, assuming the temperature (fitting) parameter either dependent or independent of the
$\kappa$-index \citep{Lazar-etal-2015, Lazar-etal-2016}. Modelling with a 
$\kappa$-independent temperature is convenient in computations, and it is therefore very much 
invoked in theoretical predictions on dispersion and stability properties of Kappa
distributed plasmas, see the reviews by \citet{Hellberg-etal-2005} and \citet{Pierrard-Lazar-2010}.
The effects of a varying $\kappa$ (including the Maxwellian limit $\kappa \to \infty$) are 
also easier to parametrize if temperature is taken constant. However, for studies 
intended to outline the effects of suprathermal populations, \citet{Lazar-etal-2015,Lazar-etal-2016} 
have shown that only the approach assuming a $\kappa$-dependent temperature may provide 
a fairly rigorous comparison of the global Kappa describing the observed distribution, with 
the cooler Maxwellian $\kappa \to \infty$ limit reproducing the core of the distribution. 
Moreover, only in this case the instabilities driven by the kinetic anisotropies
of plasma particles are systematically stimulated by the suprathermals (i.e., lowering
the values of $\kappa$), confirming the expectation that these populations should contribute
with an additional free energy \citep{Lazar-etal-2015, Lazar-etal-2016b}. 
From a different perspective including, for instance, the 
origin of Kappa distributions, there are processes responsible for the formation of 
suprathermal tails, e.g., particle energization by the wave turbulence, which may imply a 
$\kappa$-independent evolution of the velocity distributions \citep{Yoon-2014, 
Lazar-etal-2016}. In this case, the high-energy tails are counter-balanced by a 
low-velocity enhancement, so that the temperature remains indeed independent of 
parameter $\kappa$. 

A global Kappa distribution implies a reduced number of parameters and is, therefore, easy to 
manipulate in observational and theoretical analyses. However, a global Kappa 
does not always provide a good fit to the observed distributions, as we will see here 
below, e.g., the analysis in section~2 (also Figs.~\ref{f1} and \ref{f2}), which suggests that a global Kappa is just a 
simplified (or idealized) approach that artificially constrains the core and suprathermal populations 
to be described by the same parameters, e.g., density, temperature, anisotropy (not fully
justified yet, although some studies in this direction have been done by \citet{Leubner-2004, Lazar-etal-2015,
Lazar-etal-2016}). 
It may be, therefore, that fitting measurements of the observed distributions with global 
Kappa models can not provide relevant evidences to support any of the two alternatives of 
a temperature either dependent or independent of the power-index $\kappa$.

More complex models that combine distinct components are in general more
accurate in reproducing the  distributions measured in the solar wind, and may, therefore, 
unveil details of the observed distributions. For instance, the nonstreaming distributions
detected in the slow winds and at sufficiently large distances from the Sun 
\citep{Maksimovic-etal-2005} are expected to be better described by a dual model comprising a standard 
Maxwellian distribution function to reproduce the core at low energies, and a Kappa power-law 
to fit the suprathermal tails (halo component) of the distribution (see Figs.~\ref{f1} 
and \ref{f2}). In the fast winds the observed distributions become asymmetric 
(see Fig.~\ref{f3}), and to describe them we need an additional component  (called strahl) streaming 
along the magnetic field direction. Such complex models have already been  
invoked in observational analyses in order to quantify the main physical properties of these components, e.g., 
densities, relative drifts, temperatures, and even temperature anisotropies 
\citep{Maksimovic-etal-2005, Nieves-Chinchilla-etal-2008, Stverak-etal-2008, 
Vinas-etal-2010, Pierrard-etal-2016}. The halo component can be called
\textit{suprathermal} as long it is well reproduced by Kappa power-laws 
with sufficiently low values of $\kappa$, i.e., $\kappa < 10$. Otherwise, for higher 
values, i.e., $\kappa > 10$, the halo component appears to be thermalized and well 
described by a Maxwellian, but it remains a distinct component, being, in general, 
less dense but hotter than the core. In this case, the overall distribution can be described 
by a two-Maxwellians model, as a limiting $\kappa \to \infty$ case of a dual Maxwellian-Kappa 
representation. Although the existence of a halo component with $\kappa > 10$ is not so evident
from observations, a two-Maxwellians model has already been invoked in theoretical predictions 
on the dispersion and stability of space plasmas \citep{Gary-etal-1975, Gary-etal-1994} 
as well as in observational analyses \citep{Feldman-etal-1975, Pilipp-etal-1987} and in comparisons 
with global Kappa \citep{Maksimovic-etal-1997}. Note that this Maxwellian limit ($\kappa \to \infty$) of
a Kappa describing only the halo component is completely distinct from the Maxwellian fit to the core, 
and a correlation between these two is not justified in this case (as for a global Kappa modelling, 
where the Maxwellian limit may need to reproduce the core of the distribution to compare with 
the global Kappa and emphasize the effects of the suprathermals). 

When the core and suprathermal components are anisotropic, e.g., due to temperature anisotropies or 
relative drifts, a dual Maxwellian-Kappa modelling becomes complicated to deal with, even 
in numerical computations. Recently, an important progress has been made for overcoming 
this challenge in studies of the stability and relaxation of a dual core-halo plasma with 
intrinsic temperature anisotropies \citep{Lazar-etal-2014, Lazar-etal-2015b, Shaaban-etal-2016}.
Both interpretations were adopted for the anisotropic halo modelled by a bi-Kappa with 
temperatures being either dependent or independent of the power-index $\kappa$.
The results from these studies were again favorable to a Kappa model with a $\kappa$-dependent 
temperature, since only this approach provided a systematic stimulation of the instabilities
in the presence of suprathermals. 

The present paper aims to provide additional insight into these properties of Kappa distributed 
plasmas as they are revealed by observational data. As discussed above, there are two fitting 
models widely invoked in theory and observations, namely, the global Kappa and the dual 
Maxwellian-Kappa models. In section~2 we describe them mathematically and analyze qualitatively 
their accuracy in reproducing the solar wind observational data. Particularly relevant for our 
analysis is the Kappa model that better reproduces the observed distributions. This model is then 
considered for fitting to more than 100~000 velocity distributions	measured in 
the ecliptic within an wide interval of heliocentric distance between 0.3 and 4.0~AU. 
Determined as fitting parameters the temperature and the power-index $\kappa$ are subjected
to a statistical analysis that enables us to extract the relationship between them.  
The $\kappa$-dependency indicated by the observations for the temperature of Kappa populations 
is used in section~3 for a kinetic analysis of the electromagnetic instabilities 
driven by the temperature anisotropy. These instabilities can explain the enhanced fluctuations 
observed in the solar wind, see the review by \cite{Zimbardo-etal-2010} and references therein.
The results of our study are summarized and discussed in section~4.

\section{Refined models from the observations}

\subsection{Global Kappa vs. dual Maxwellian-Kappa}

\begin{figure}[h] \centering
\includegraphics[width=75mm]{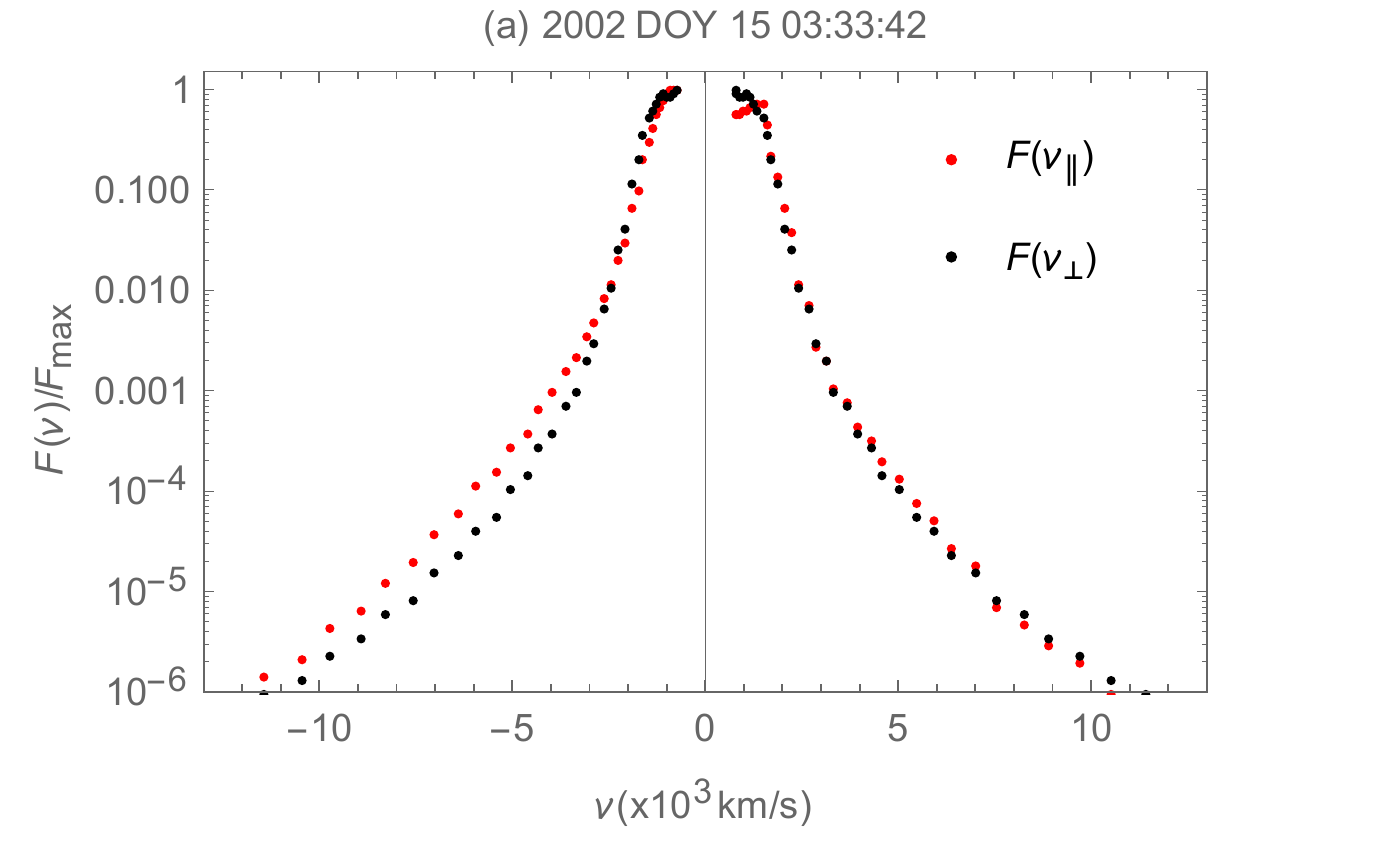}\\
\includegraphics[width=75mm]{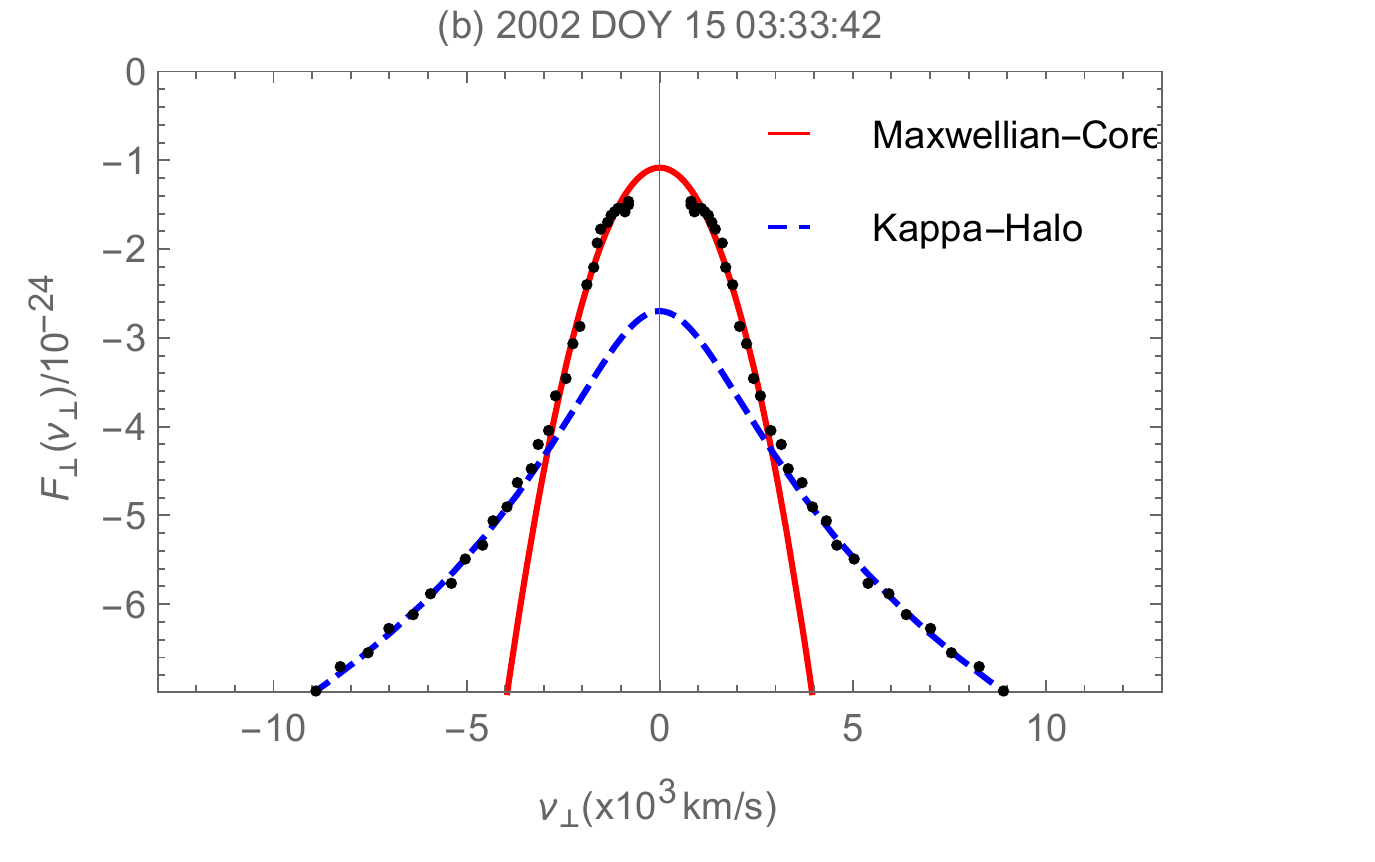}\\
\includegraphics[width=75mm]{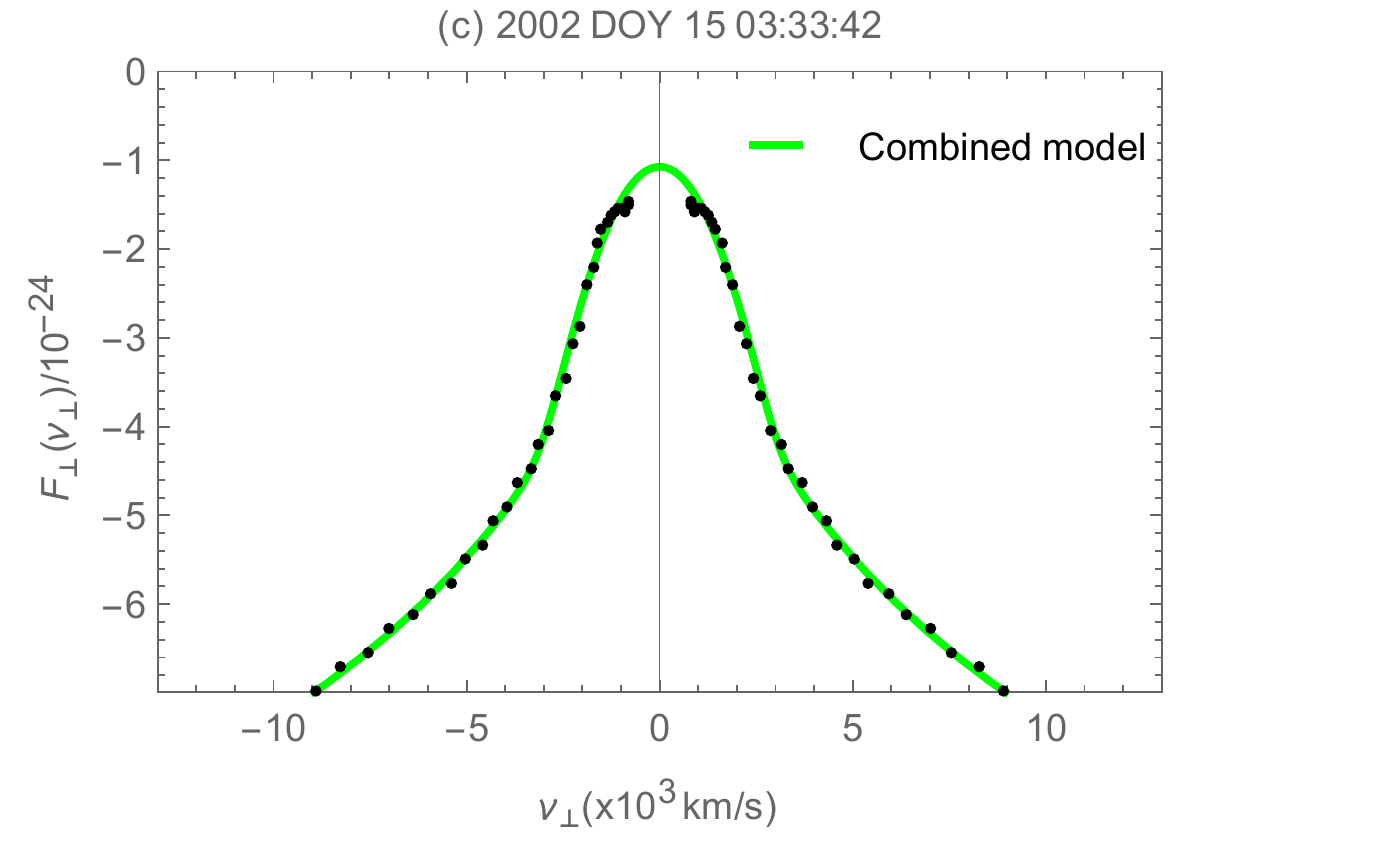}\\
\includegraphics[width=75mm]{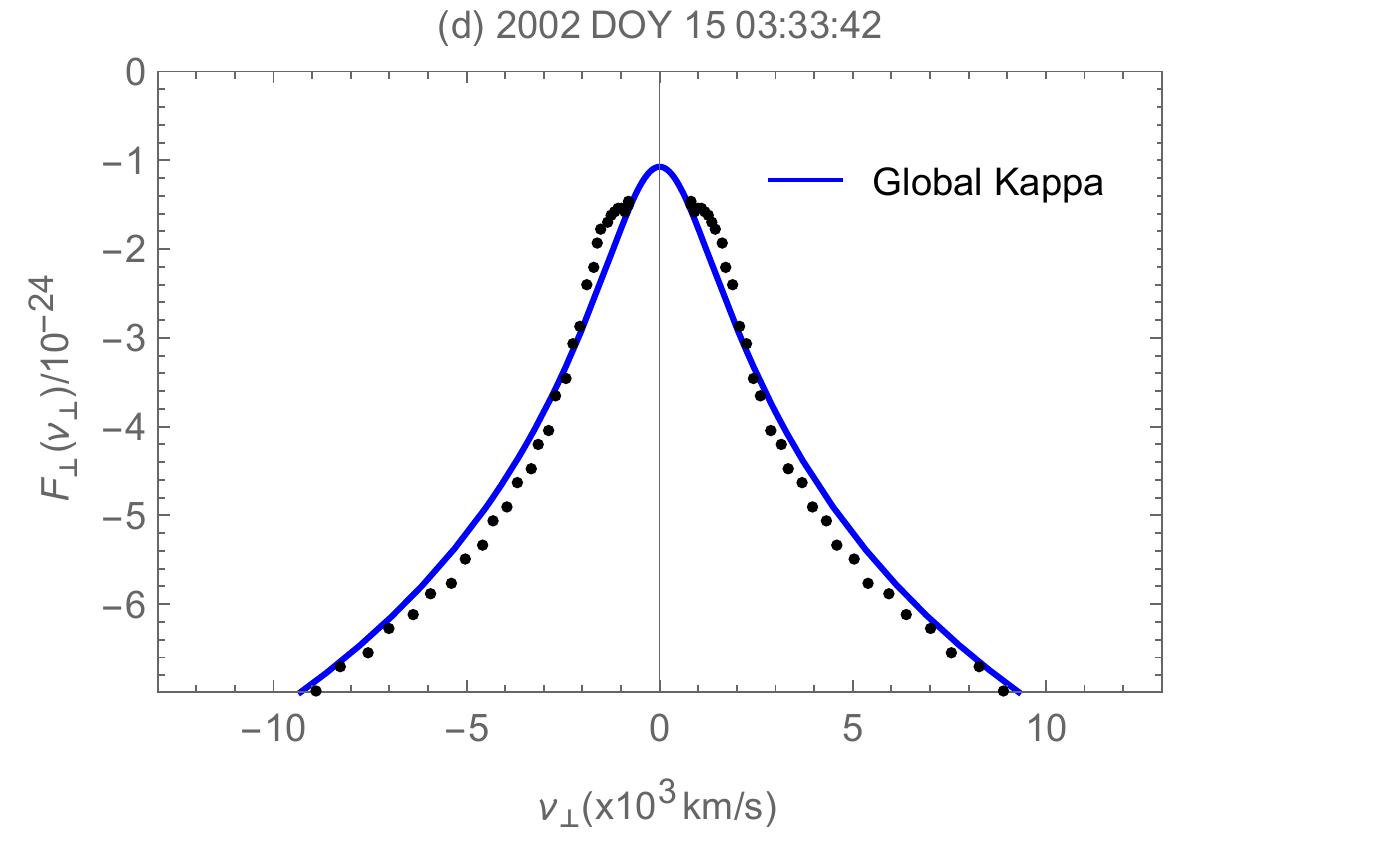}
    \caption{Nonstreaming electron velocity distribution $F$~[cm$^-6$ s$63$] measured by \emph{Ulysses} in
    the slow wind in 2002 on DOY 15 03:33:42: (a) parallel and perpendicular cuts; (b) 
		Maxwellian fit to the core and Kappa fit to the halo suprathermal tails; (c) 
		combined (resultant) Maxwellian-Kappa model fitting the entire distribution; (d)
		global Kappa model fitting the entire distribution.} \label{f1}%
\end{figure}

\begin{figure}[h] \centering
\includegraphics[width=75mm]{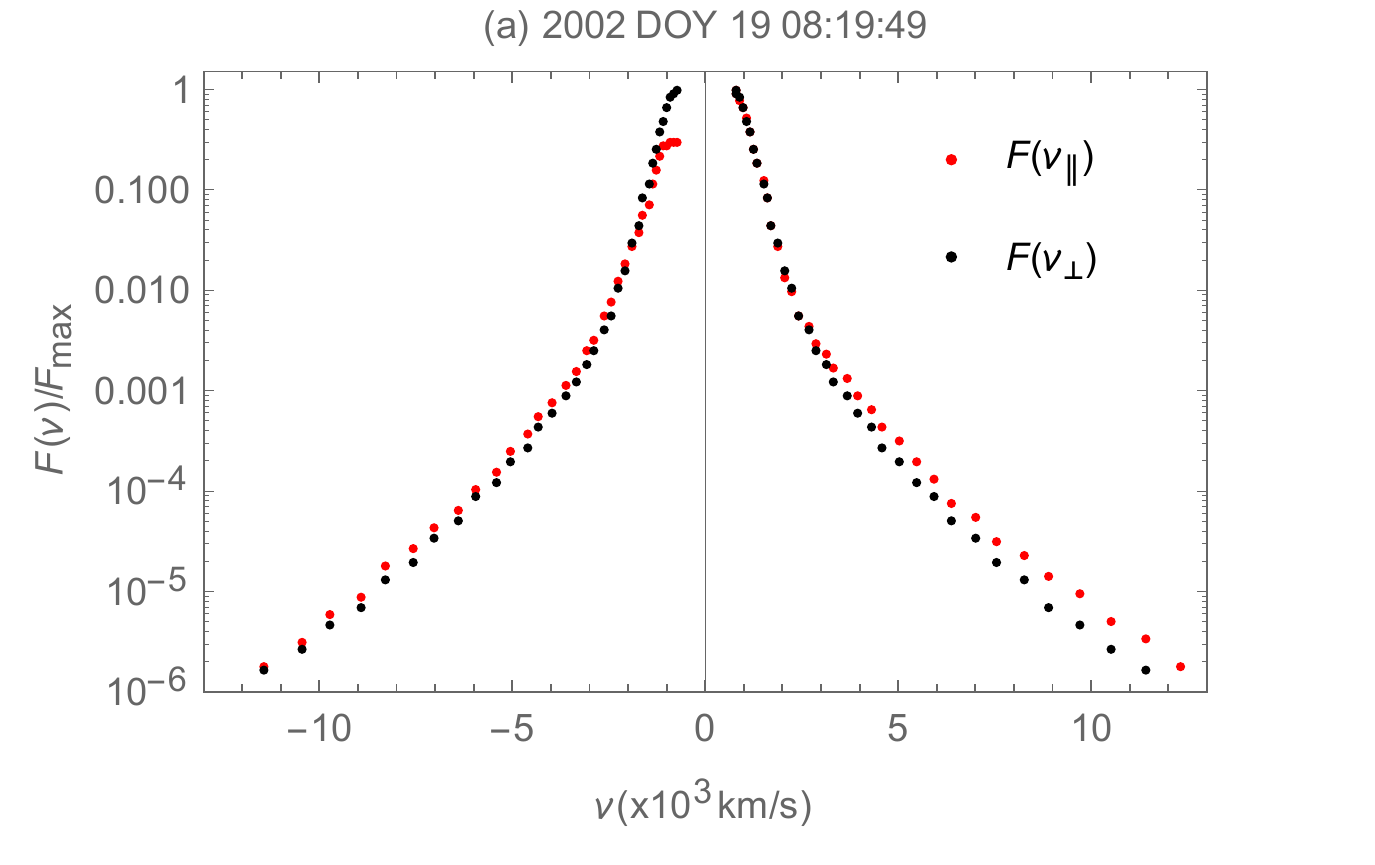}\\
\includegraphics[width=75mm]{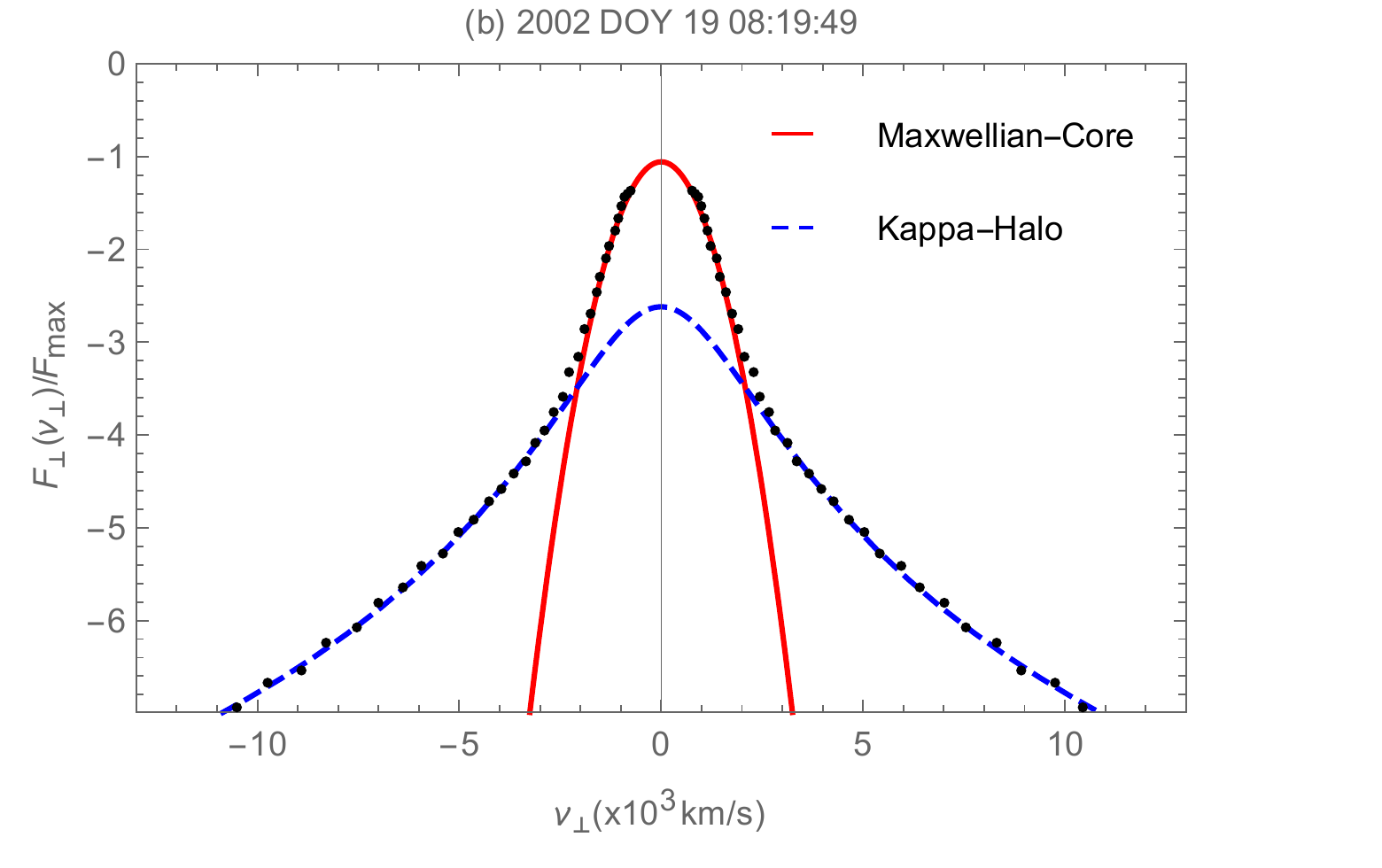}\\
\includegraphics[width=75mm]{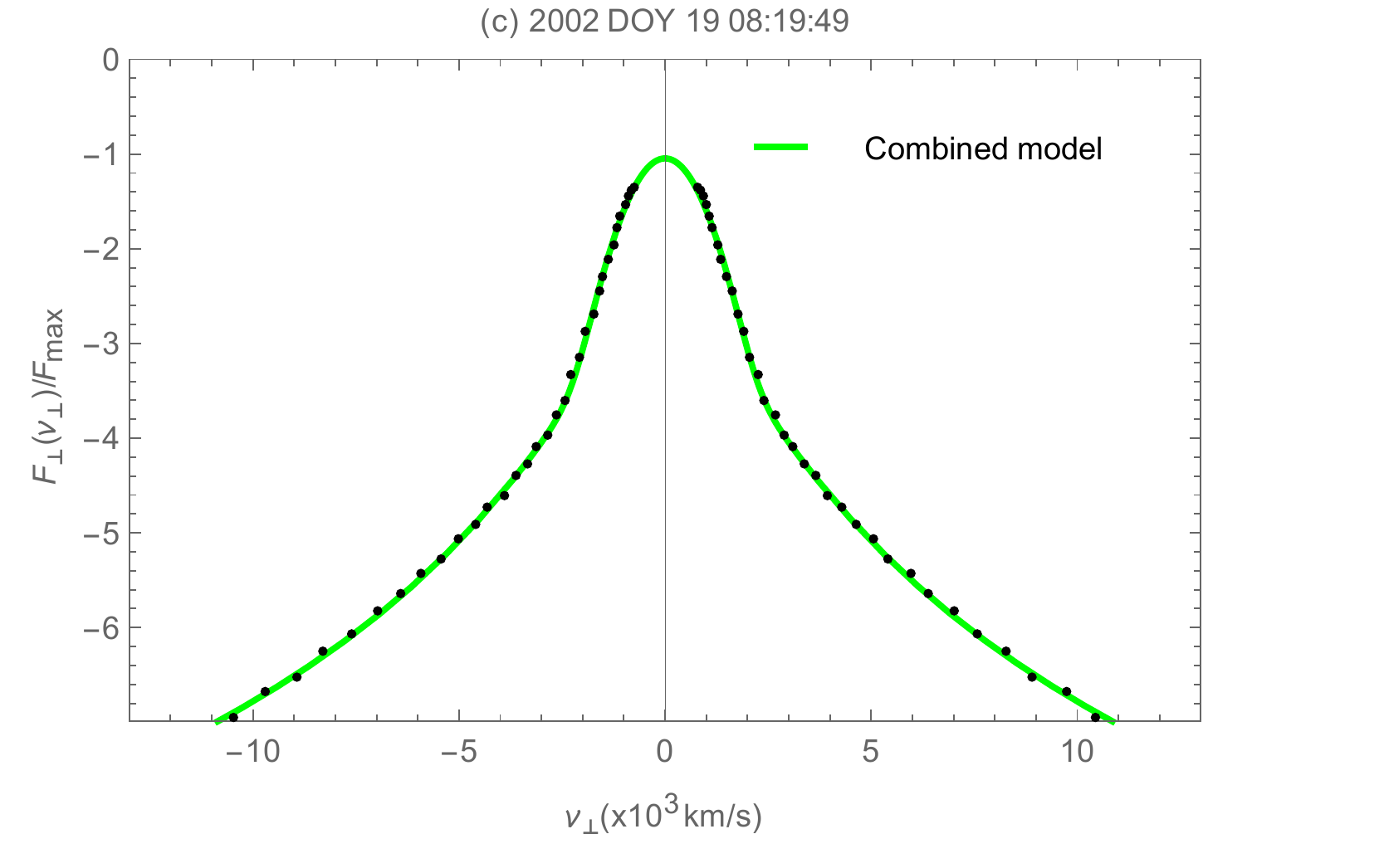}\\
\includegraphics[width=75mm]{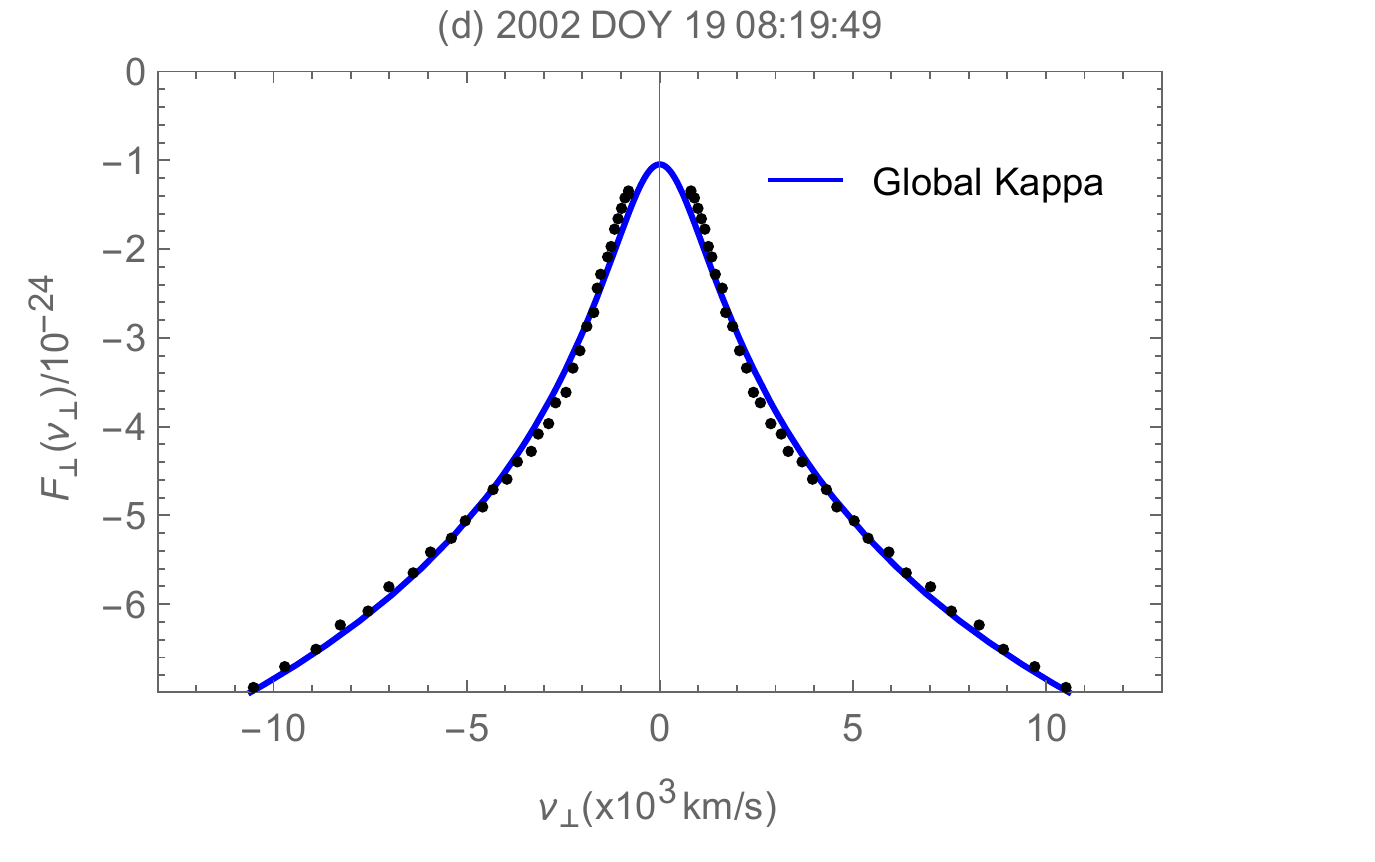}
    \caption{Nonstreaming electron velocity distribution $F$~[cm$^-6$ s$63$] measured by \emph{Ulysses} in
    the slow wind in 2002 on DOY 19 08:19:49: (a) parallel and perpendicular cuts; (b) 
		Maxwellian fit to the core and Kappa fit to the halo suprathermal tails; (c) 
		combined (resultant) Maxwellian-Kappa model fitting the entire distribution; (d)
		global Kappa model fitting the entire distribution.} \label{f2}%
\end{figure}

Fluxes of plasma particles measured in-situ in the solar wind are transformed
into the frame of bulk flow, such that, in the absence of any differential streaming, 
the velocity distributions are almost symmetric in the Sun-ward and anti-Sun-ward directions, 
without indications of any additional strahl (or streaming) component (see examples in 
Figs.~\ref{f1} and \ref{f2}). One intriguing feature that makes them impossible 
to describe with standard Maxwellian models are the enhanced high-energy tails, known
as suprathermal tails. Instead, the Kappa (Lorentzian) power-laws are used to reproduce 
the suprathermal tails of the observed distributions, and this can be done in two distinct
ways. Thus, a Kappa power-law (subscript $\kappa$) is used either as a global model (subscript $GK$) 
to fit the entire distribution 
\begin{equation} f_{GK}(v_{\parallel}, v_{\perp}) = n f_{\kappa}(v_{\parallel}, v_{\perp}), 
\label{e1} \end{equation} 
or to describe only the suprathermal halo (subscript $h$) component 
\begin{equation} f_h(v_{\parallel}, v_{\perp}) = f_{\kappa}(v_{\parallel}, v_{\perp}).
\label{e2} \end{equation} 
In the second case the core (subscript $c$) fitting is a Maxwellian (subscript $M$)
\begin{equation} f_c (v_{\parallel}, v_{\perp}) = f_M (v_{\parallel}, v_{\perp}),
\label{e3} \end{equation} 
and the resulting model to describe the entire distribution $f$ is a dual Maxwellian-Kappa
(subscript $MK$)
\begin{equation} f_{MK}(v_{\parallel}, v_{\perp}) = n_c f_{c}(v_{\parallel}, v_{\perp}) + 
n_h f_h(v_{\parallel}, v_{\perp}), \label{e4} \end{equation} 
Since kinetic modelling deals with homogeneous plasma systems in general, in the above $n$ 
is the total number density of plasma particles (of given species), and 
$n_c$ and $n_h$ are the number densities of the core and halo components, respectively.
Moreover, in magnetized plasmas from space the velocity distributions are in general gyrotropic (without 
a major anisotropy in the plane transverse to the magnetic field) allowing us to describe them in
polar coordinates $(v_x, v_y, v_z)=(v_{\perp} \cos \phi, v_{\perp} \sin \phi, v_{\parallel})$
with $\parallel$ and $\perp$ denoting directions relative to the stationary magnetic field.
Gyrotropic distribution functions reduce the analysis to only two variables in velocity space 
and enable us to quantify the principal anisotropies reported by the observations, e.g., 
the bi-axis temperature anisotropies $T_\perp \ne T_\parallel$ combined or not with 
the field-aligned streams (also known as strahls). 

Relevant for the nonstreaming distributions observed in the solar wind is the
well-known bi-Kappa distribution function \citep{Summers-Thorne-1991, Lazar-etal-2015}
\begin{eqnarray}
f_{\kappa} (v_{\parallel}, v_{\perp}) & = & {1 \over \pi^{3/2}
\theta_{\perp}^2 \theta_{\parallel}} \, {\Gamma(\kappa +1) \over
\kappa^{3/2} \Gamma(\kappa -1/2)} \left(1 + {v_{\parallel}^2\over
\kappa \theta_{\parallel}^2 } + {v_{\perp}^2\over \kappa
\theta_{\perp}^2 }\right)^{-\kappa-1}
\nonumber \\
& = & \left[m \over \pi k_B (2\kappa-3)\right]^{3/2}
{1 \over T_\perp \sqrt{T_\parallel}}
{\Gamma(\kappa +1) \over \Gamma(\kappa -1/2)}
\nonumber\\
& \times & \left[1+ {m \over k_B (2\kappa-3)} \left(
{v_{\parallel}^2\over T_{\parallel} } + {v_{\perp}^2\over
T_{\perp} }\right) \right]^{-\kappa-1}, \label{e5}
\end{eqnarray}
where $v_{\|}$ and $v_{\perp}$ denote particle velocity parallel and
perpendicular w.r.t.\ a large-scale magnetic field, $m$ is particle mass, $k_B$ the Boltzmann constant,
$\Gamma$ is the Gamma function, the power-index $\kappa \in (3/2,\infty]$, $T_{\|,\perp}$
and $\theta_{\|,\perp}$ the corresponding temperatures and thermal
velocities, which are related by
\begin{equation} T_{\parallel} \equiv {2 m \over 2 k_B} \int d{\bf v}
v_\parallel^2 f_{\kappa}(v_{\parallel}, v_{\perp}) = {\kappa
\over \kappa -3/2} {m \theta_{\parallel}^2 \over 2k_B}, \label{e6} \end{equation}
\begin{equation} T_{\perp} \equiv {m \over 2 k_B} \int d{\bf v} v_\perp^2
f_{\kappa}(v_{\parallel}, v_{\perp}) = {\kappa \over \kappa -3/2} {m \theta_{\perp}^2
\over 2 k_B}. \label{e7} \end{equation}
%
%
%
This bi-Kappa model is invoked in both fitting techniques, either as a global model to 
describe the entire distribution \citep{Vasyliunas-1968, Christon-etal-1989, Collier-etal-1996, 
Maksimovic-etal-1997}, or to fit partially, only the suprathermal tails of the distribution
\citep{Maksimovic-etal-2005, Nieves-Chinchilla-etal-2008, Stverak-etal-2008, 
Pierrard-etal-2016}.


The low-energy core is well reproduced by a bi-Maxwellian (e.g., in \cite{Pilipp-etal-1987})
\begin{equation} f_M (v_{\parallel}, v_{\perp}) = {1 \over \pi^{3/2}
w_{\parallel} w_{\perp}^2} \, \exp \left(- {v_{\parallel}^2\over
w_{\parallel}^2 }- {v_{\perp}^2\over w_{\perp}^2 }\right),
\label{e8} \end{equation}
where $w_{\parallel,\perp}$ are thermal velocities defined by the 
temperatures as moments of second order
\begin{equation} T_{\parallel} \equiv {2 m \over 2 k_B} \int d{\bf v}
v_\parallel^2 f_M(v_{\parallel}, v_{\perp}) = {m w_\parallel^2 \over 2k_B},
\label{e9} \end{equation}
\begin{equation} T_{\perp} \equiv { m \over 2 k_B}\int d{\bf v} v_\perp^2
f_M(v_{\parallel}, v_{\perp}) = {m w_\perp^2 \over 2k_B}. \label{e10} \end{equation}

The so-called global Kappa approach from Eq.~(\ref{e1}) implies a reduced number of parameters 
and is, therefore, preferred in computations, while a more complex, dual Maxwellian-Kappa 
as defined in Eq.~(\ref{e4}) is expected to reproduce better the observations, although direct comparisons
with global Kappa fittings are not reported yet (at least to our knowledge). In Figs.~\ref{f1} 
and \ref{f2} we compare these two fitting models as obtained for the electron velocity distributions  
measured by Ulysses in January 2002 (dotted lines), two events on DOY 15 03:33:42 and DOY 19 08:19:49,
respectively. These distributions are chosen to be almost isotropic, i.e., with insignificant 
differences between parallel (red dots) and perpendicular (black dots) cuts, see top panels a, 
such that fittings to 
the perpendicular cut are representative for the entire distribution. 

\begin{figure}[h] \centering
\includegraphics[width=75mm]{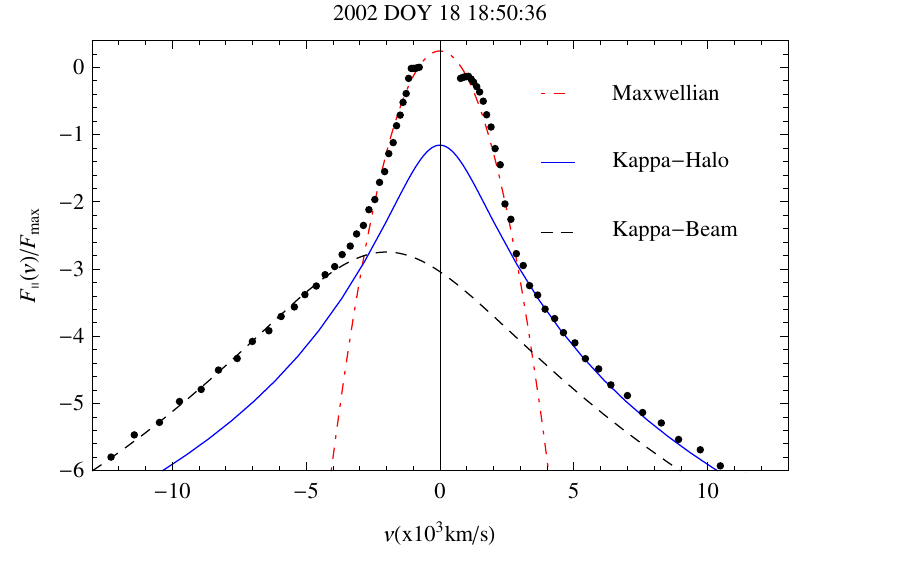}\\
\includegraphics[width=75mm]{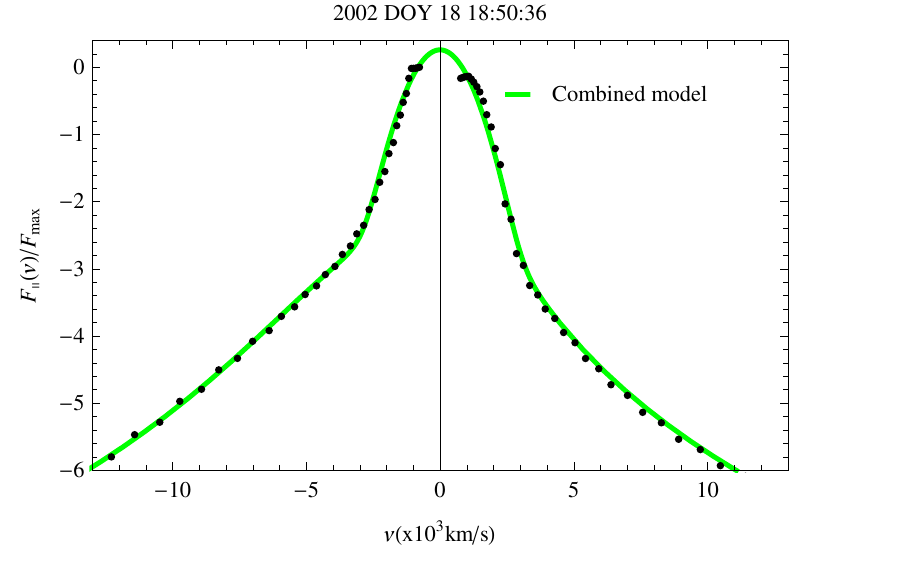}\\
    \caption{Asymmetric electron distribution $F$~[cm$^-6$ s$63$] measured by \emph{Ulysses} in 2002 
		DOY 18 18:50:36: (a) Maxwellian fit to the core, Kappa fit the halo suprathermal tails, and a
		drifting Kappa fit to the streaming (strahl) component; (b) combined (resultant) model 
		fitting the entire distribution.} \label{f3}%
\end{figure}

The global Kappa fit is displayed in the bottom panels (solid blue lines), and the 
dual Maxwellian-Kappa fit is explicitly shown in the middle
panels: in the second panel from top we have the core (solid red lines) and halo (blue dashed lines) 
fits and these two are then combined as a dual Maxwellian-Kappa fit in the third panel (solid green lines). 
Now, a direct comparison of these two fitting 
models becomes possible just by analyzing the last two bottom panels in Figs.~\ref{f1} and \ref{f2}.
It is obvious that in both events the observed distribution is better reproduced
by a dual Maxwellian-Kappa approach than a global Kappa. Note that in Fig.~\ref{f2} a global 
Kappa provides a better fit to the observed distribution than in Fig.~\ref{f1}, but in both 
these two cases the global Kappa fits are visibly less accurate than those obtained with a dual 
Maxwellian-Kappa approach. 
In this case the effects of suprathermals may simply be outlined by a direct comparison between the dual model
$f_{MK}$ and the Maxwellian fit $f_M$ to the core, and there is no need to consider the Maxwellian limit 
$\kappa \to \infty$ of the Kappa component. 

The fact that a dual model is more accurate 
may simply suggest the existence of two distinct components, namely, a low-energy 
core population and a suprathermal Kappa-distributed halo. There is probably no other possibility
to make distinction between these components, which may have different origins in the solar corona
\citep{Pierrard-etal-1999} or in the solar wind \citep{Maksimovic-etal-2005}. On the other 
hand, these two components are expected to be strongly correlated, especially by the collective effects 
of plasma particles which should tend to diminish differences between them. However, this is not 
always confirmed or evident from observations, which show, for instance, an electron halo 
enhancing with the radial distance from the Sun on the expense of the more anisotropic strahl
which is diminished in almost the same measure \citep{Maksimovic-etal-2005}. In this case an 
explanation may be offered by the instabilities and amplified fluctuations selfgenerated by the 
strahl, which can isotropize the strahl, scattering the particles and thus feeding the more isotropic
suprathermal halo. 
A dual core-halo model is relevant not only in the slow winds when the flowing 
(bulk) speed is sufficiently low (i.e., $V_{\rm SW} < 360$ km s$^{-1}$), and the relative density
of the strahl is negligibly small $n_s < n_h < n_c$ \citep{Lazar-etal-2015b}, but according to \cite{Maksimovic-etal-2005} 
it may also describe any slow or fast wind conditions at sufficiently large 
heliocentric distances, e.g., beyond 1~AU where the strahl component is markedly
reduced. A dual Maxwellian-Kappa remains a major component in the fast winds, when the velocity 
distribution exhibits an additional streaming component (called strahl). In Fig.~\ref{f3} 
we display an example of velocity distribution of electrons measured during a fast solar wind 
by Ulysses in January 2002. In this case the best fitting model contains three components: 
a Maxwellian core, a Kappa halo, and a drifting-Kappa strahl (or beam of particles).

%

\begin{figure}[ht]
    \includegraphics[width=75mm]{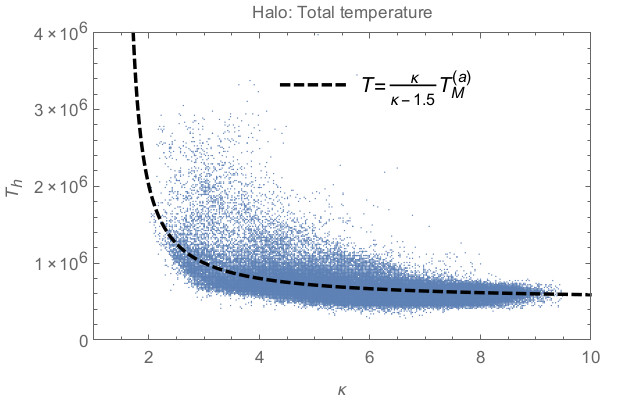}\\
    \includegraphics[width=75mm]{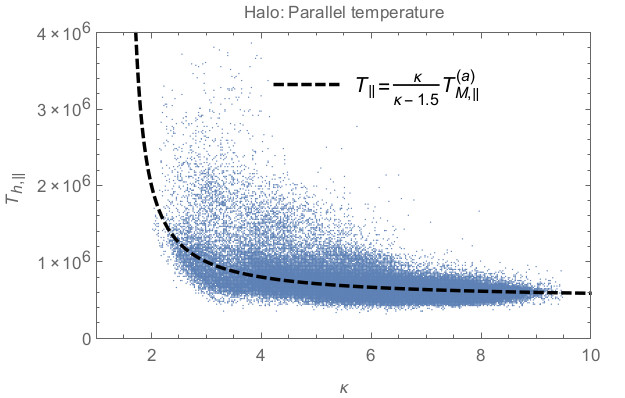}\\
    \includegraphics[width=75mm]{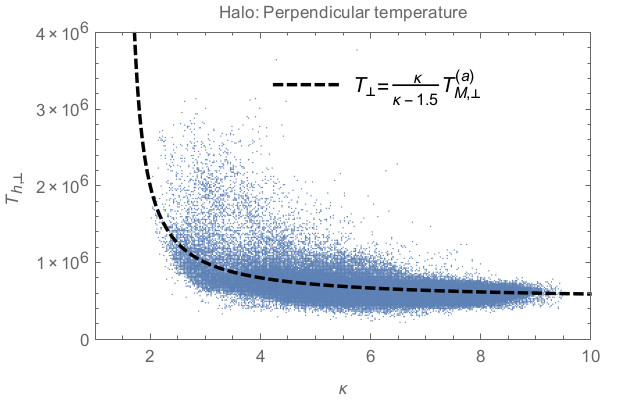}
\caption{Scatter plots (blue dots) of the halo temperature $T$~[K] vs. $\kappa$-index, 
measured in the ecliptic (0.3~AU $\leqslant R < $ 4~AU): total temperature (top), 
and the components, parallel temperature (middle), and perpendicular temperature (bottom).
The temperatures predicted by the Eqs.~(\ref{e15})--(\ref{e17}) as functions of 
$\kappa$ parameter are overplotted with dashed lines.} \label{f4}
\end{figure}

\subsection{The Kappa (halo) temperature from observations}

If a $\kappa$-dependency exists for the temperature of Kappa population, then this 
must be revealed by the observational data. In order to be relevant the Kappa model should 
accurately reproduce the measured distribution, and here above we have demonstrated that
this condition is satisfied by a dual Maxwellian-Kappa model, with Kappa distribution
function fitting only the suprathermal (halo) tails.

\citet{Stverak-etal-2008} have used the same dual Maxwellian-Kappa model from Eq.~(\ref{e4}) 
to determine the principal moments of the electron distributions measured in 
the solar wind. The dual model was fitted to more than 100~000 velocity distributions
measured by three different spacecraft missions (Helios 1, Cluster II, and Ulysses) 
in the ecliptic at different heliocentric distances in the interval 0.3--3.95~AUs.
Details about the electron analyzers and the methods of correction and reconstruction of 
the 3D velocity distributions are given in \citet{Stverak-etal-2008} and some references therein. 
Here we make use of the same data set, and investigate the temperature parameters determined
for the Kappa (halo) populations. Fig.~\ref{f4} displays scatter plots (with blue dots) of
the temperatures determined for the electron halo (subscript $h$) \textit{vs.} the corresponding 
values of the power-index $\kappa$: total temperature ($T_h$) in the top panel, parallel 
($T_{h,\parallel}$) and perpendicular ($T_{h,\perp}$) temperatures in the middle and bottom 
panels, respectively. One dimensional fits provide temperature components $T_{h,\parallel}$ 
and $T_{h,\perp}$, enabling us to calculate $T_h = T_{h,\parallel}/3 + 2 T_{h,\perp}/3$.

For each panel in Fig.~\ref{f4} the spread of data points is very similar and we can analyze 
them generically. The data points concentrate along a curve which clearly shows that temperature 
is not constant but varies as a function of $\kappa$. Thus, the 
temperature of Kappa population decreases with increasing power-index $\kappa$, with the allure of 
an asymptotic decrease towards the Maxwellian limit $\kappa \to \infty$. In this 
limit of a very large $\kappa \to \infty$ the bi-Kappa reproducing the halo 
components reduces to a bi-Maxwellian (similar to that reproducing the core 
component). However, \citet{Lazar-etal-2016} have shown that any Kappa 
distribution function admits two distinct Maxwellian limits:
\begin{description}
	\item[\textbf{(a)}] a cooler Maxwellian limit corresponding to the approach with a $\kappa$-dependent temperature
\begin{equation} f_{M}^{(a)}=\lim_{\kappa \to \infty} f_{\kappa} (v_{\parallel}, v_{\perp}) = {1 \over \pi^{3/2}
\theta_{\parallel} \theta_{\perp}^2} \exp \left(- {v_{\parallel}^2\over
\theta_{\parallel}^2 } - {v_{\perp}^2\over \theta_{\perp}^2 }\right), \label{e11} \end{equation}
with thermal velocities $\theta_{\parallel, \perp}$ determined in this case by lower temperatures 
\begin{equation} T_{M,\parallel}^{(a)} = \lim_{\kappa \to \infty} T_{\parallel} = 
{m\theta_{\parallel}^2\over 2k_B} \leqslant {\kappa \over \kappa -3/2} {m\theta_{\parallel}^2
\over 2k_B} = T_{\parallel}, \label{e12} \end{equation}
\begin{equation} T_{M,\perp}^{(a)} = \lim_{\kappa \to \infty} T_{\perp} = {m\theta_{\perp}^2 
\over 2k_B} \leqslant {\kappa \over \kappa -3/2} {m\theta_{\perp}^2 \over 2k_B} = T_\perp. \label{e13} \end{equation}
	\item[\textbf{(b)}] the second limit corresponds to the approach with a constant temperature, not dependent on $\kappa$:
\begin{eqnarray} f_{M}^{(b)} & = & \lim_{\kappa \to \infty} f_{\kappa} (v_{\parallel}, v_{\perp}) 
\nonumber \\ & = &\left({m \over 2 \pi k_B}\right)^{3/2}
{1\over \sqrt{T_{\parallel}} T_{\perp}} \exp \left[- {m \over 2k_B}\left({v_{\parallel}^2\over
T_{\parallel}} + {v_{\perp}^2\over T_{\perp}}\right)\right], \label{e14} \end{eqnarray}
where $T_{M,\parallel}^{(b)} = T_{\parallel}$ and $T_{M,\perp}^{(b)} = T_{\perp}$.
\end{description}
\begin{figure}[h]
		\includegraphics[width=75mm]{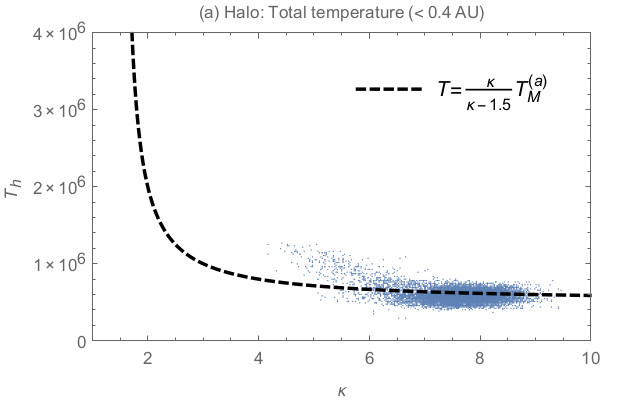}\\
		\includegraphics[width=75mm]{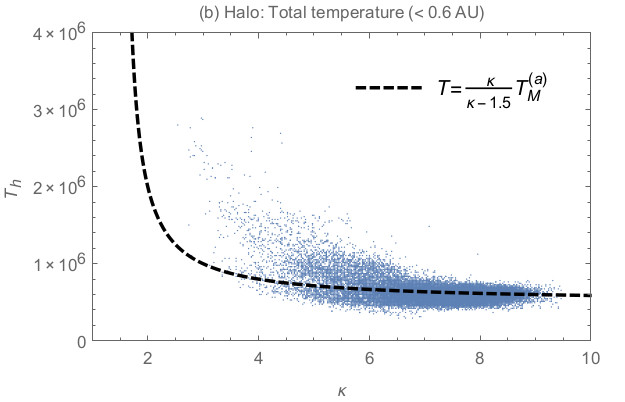}\\
		\includegraphics[width=75mm]{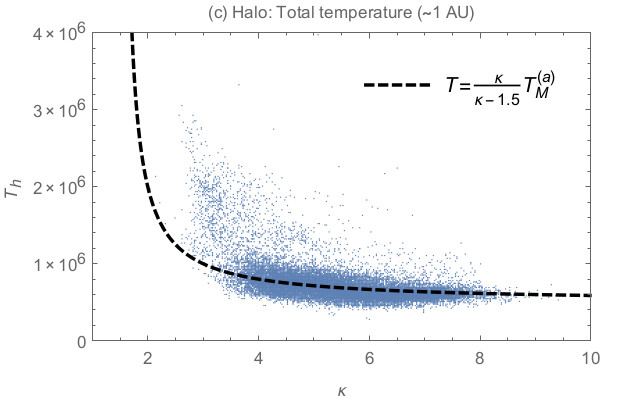}\\
		\includegraphics[width=75mm]{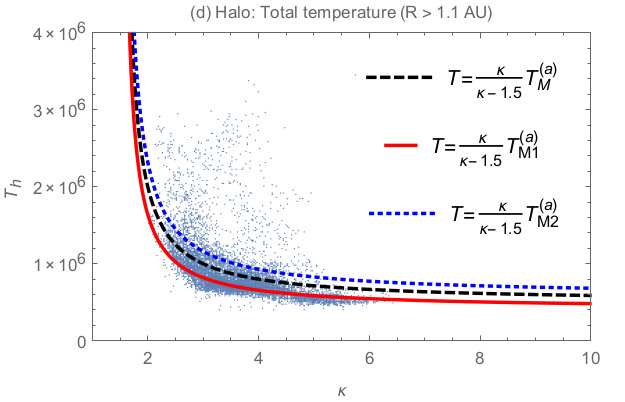}
\caption{Scatter plots (blue dots) of the halo (total) temperature $T$~[K] vs. $\kappa$-index, 
measured in different intervals of heliocentric distances $R$ (selected from the data 
set in Fig.~\ref{f4}): (a) $ 0.3< R < 0.4$~AU, (b) $< 0.6$~AU, (c) $\sim 1$~AU (0.9~AU 
$\leqslant R \leqslant$ 1.1~AU), (d) $R > 1.1$~AU. Fitting curves are explained in the text.} \label{f5}
\end{figure}

The observations indicate a $\kappa$-dependent temperature, when the components $T_{\parallel, \perp}$ 
can be defined in terms of their Maxwellian limits $T_{M,\parallel, \perp}^{(a)}$ as follows
\begin{equation}  T_{\parallel} = {\kappa \over \kappa - 3/2} {m\theta_{\parallel}^2
\over 2k_B} = {\kappa \over \kappa -1.5} T_{M,\parallel}^{(a)}, \label{e15} \end{equation}
\begin{equation}  T_\perp = {\kappa \over \kappa - 3/2} {m\theta_{\perp}^2 \over 2k_B} = 
{\kappa \over \kappa -1.5} T_{M,\perp}^{(a)}. \label{e16} \end{equation}
and a similar relation is obtained for the total temperature
\begin{equation}  T= {\kappa \over \kappa - 3/2} T_{M}^{(a)}. \label{e17} \end{equation}
Using these relations to fit to the observations, see the dashed curves in Fig.~\ref{f4}, 
we find for the Maxwellian limits $T_{M}^{(a)} \simeq T_{M,\parallel}^{(a)} \simeq 
T_{M,\perp}^{(a)} = 5 \times 10^5$~K. We should point out that the accumulation of data 
aligns with these dashed curves very well, which confirms the validity of $\kappa$-dependency
found in Eqs.~(\ref{e15})--(\ref{e17}). However, the power-index $\kappa$ describing
the electron halo population in the solar wind does not take very large values, being
limited to $\kappa < 10$. Therefore, for a rigorous interpretation of the solar wind 
electrons, a two-Maxwellian model seems to be not justified, but can be invoked as a 
limiting case of a dual Maxwellian-Kappa in order to depict and investigate the effects 
of the suprathermal populations. 

Both  parameters $T$ and $\kappa$ analyzed here characterize the halo (Kappa) electron component 
of the solar wind plasma, and should therefore undergo a variation with the radial (heliocentric) 
distance from the Sun, being naturally conditioned by the solar wind evolution during its 
expansion in space. Thus, recent studies of the same set of data \citep{Pierrard-etal-2016} have 
shown significant differences between the core and halo temperatures, and implicitly between 
their evolutions with heliocentric distance: the core temperature decreases (with a tendency 
of stabilizing after 2~AU) while the halo temperature increases with radial distance 
from the Sun (with an apparent saturation at about 3~AU). Normally, the temperature should indeed
cool down with the distance from the Sun, as the core temperature is confirming. The increase of 
the halo temperature is somehow unusual, although it is in perfect agreement with the evolution
of the other parameters reported by different observations \citep{Maksimovic-etal-2005, 
Pierrard-etal-2016}, namely, the increase of halo population (i.e, the increase of its relative 
number density $n_h/n_c$) at the same time with a decrease of $\kappa$ with radial distance.
The halo population seems to be enhanced on the expense of the strahl population, which 
is eventually scattered and isotropized by the selfgenerated instabilities (collisions are
inefficient at large distances in the solar wind). 
At this point it becomes therefore interesting to check whether such kind of processes 
may have an influence on the $\kappa$-dependency shown by the halo temperature in the observations. 

Fig.~\ref{f5} is intended to unveil the existence of these influences if they exist, by 
presenting in detail a radial profile of the halo temperature as a function of the power-index 
$\kappa$. We made a selection of the data from different heliocentric distances, and plot in 
four panels with the distance increasing from top panel (a) to bottom panel (d) as follows: 
(a) $ 0.3< R < 0.4$~AU, (b) $< 0.6$~AU, (c) $\sim 1$~AU (0.9~AU $\leqslant R \leqslant$ 1.1~AU), 
(d) $R > 1.1$~AU. At small radial distances $ 0.3< R < 0.4$~AU the data accumulates and aligns
very well to the curve defined by Eq.~(\ref{e17}) for a $\kappa$-variation of the halo temperature.
With increasing the distance the data points spread above and below this fitting curve, most probably
as an effect of some mechanisms at work in the solar wind expansion, like cooling, magnetic 
focusing, and particles scattering by the fluctuations, etc. However, up to~1AU, the data 
on average aligns well on the $\kappa$-dependency in Eq.~(\ref{e17}) with the same asymptotic
(Maxwellian) limit $T_{M2}^{(a)} \simeq 5 \times 10^5$~K. We cannot claim the same thing for the data collected
beyond 1~AU, which on average seems to align to the same $\kappa$-dependency law as the one given
in Eq.~(\ref{e17}), but with a lower Maxwellian (asymptotic) limit $T_{M1}^{(a)} \simeq 4.1 \times 
10^5$~K, see the the solid (red) line in panel (d). Moreover the spread of data above the dashed 
line suggests the existence of other hotter populations which may fit to another similar 
law but with a higher Maxwellian (asymptotic) limit $T_{M2}^{(a)} \simeq 5.8 \times 
10^5$~K, see the dotted (blue) line in panel (d).
In order to explain these distinct populations indicated by different fitting curves (and leading 
to a spread of data along the average fitting) plausibly would be to invoke the different origins
of these populations. For instance, scattering of the beaming strahl by the selfgenerated 
instabilities may provide hotter Kappa populations \citep{Maksimovic-etal-2005}, as also discussed 
here above.
We can conclude stating that the observations in the solar wind show clear evidences 
that support a $\kappa$-dependency of the temperature of Kappa distributed populations, as 
the one explicitly shown above in Eqs.~(\ref{e15})-(\ref{e17}).

\begin{figure}[t]
  \centering
     \includegraphics[width=0.4\textwidth]{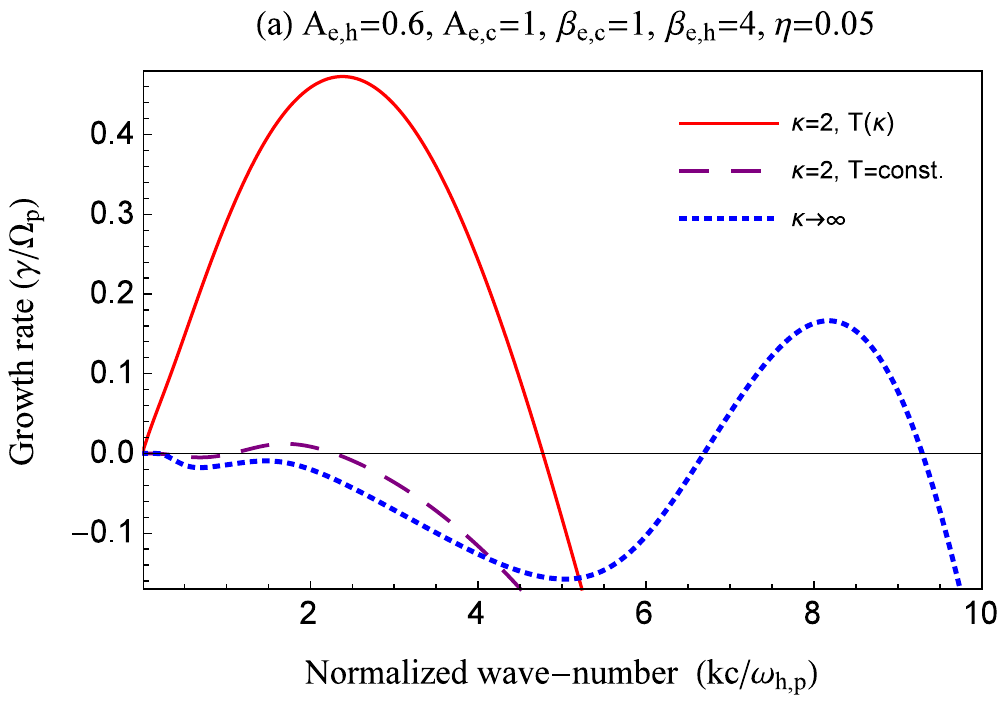}
     \includegraphics[width=0.4\textwidth]{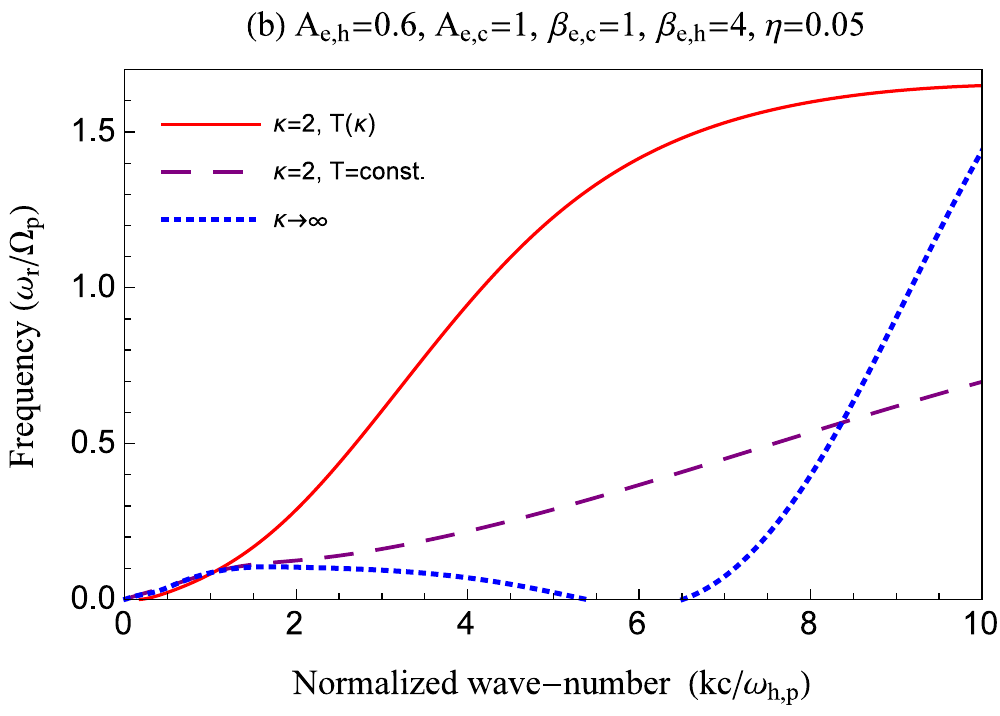}
      \caption{EFHI: growth rates (top) and real frequencies (bottom) derived for a dual model comprising 
			a Maxwellian core and a bi-Kappa halo ($\kappa =2$) with $\kappa-$dependent temperatures (solid lines),
			$\kappa-$independent temperatures (dashed lines), or their bi-Maxwellian limit ($\kappa \to \infty$,
			dotted-lines). The plasma parameters for electrons are given explicitly in each panel and for protons 
			are $A_{p,c}=A_{p,h}=1$, $\beta_{p,c}=1$, $\beta_{p,h}=4$.}\label{f6}
  \end{figure}
\begin{figure}[t] \centering
\includegraphics[width=0.4\textwidth]{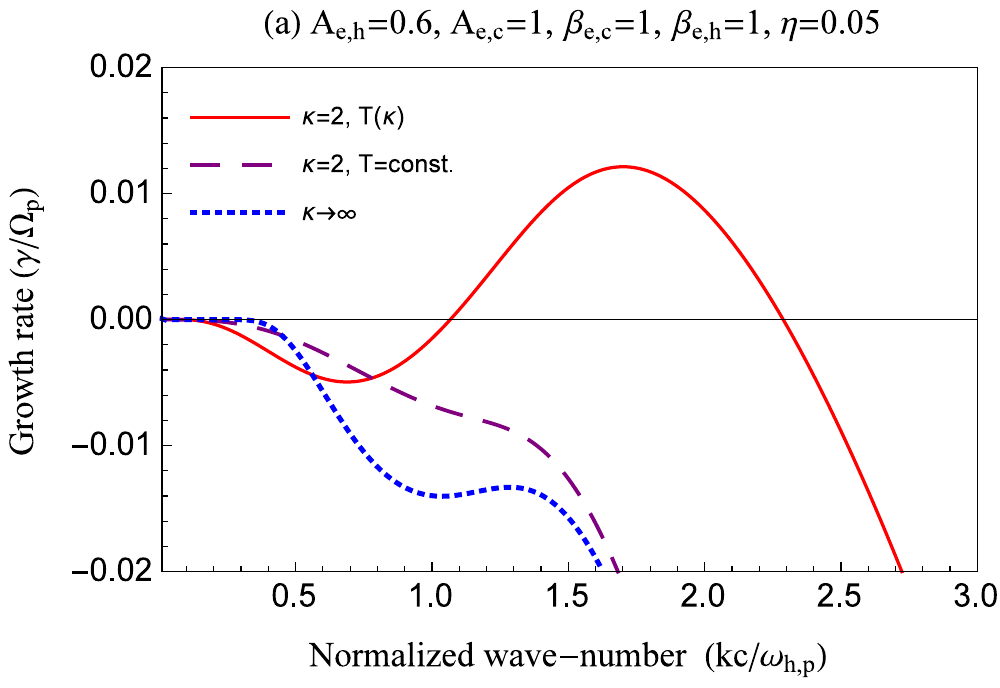}
\includegraphics[width=0.4\textwidth]{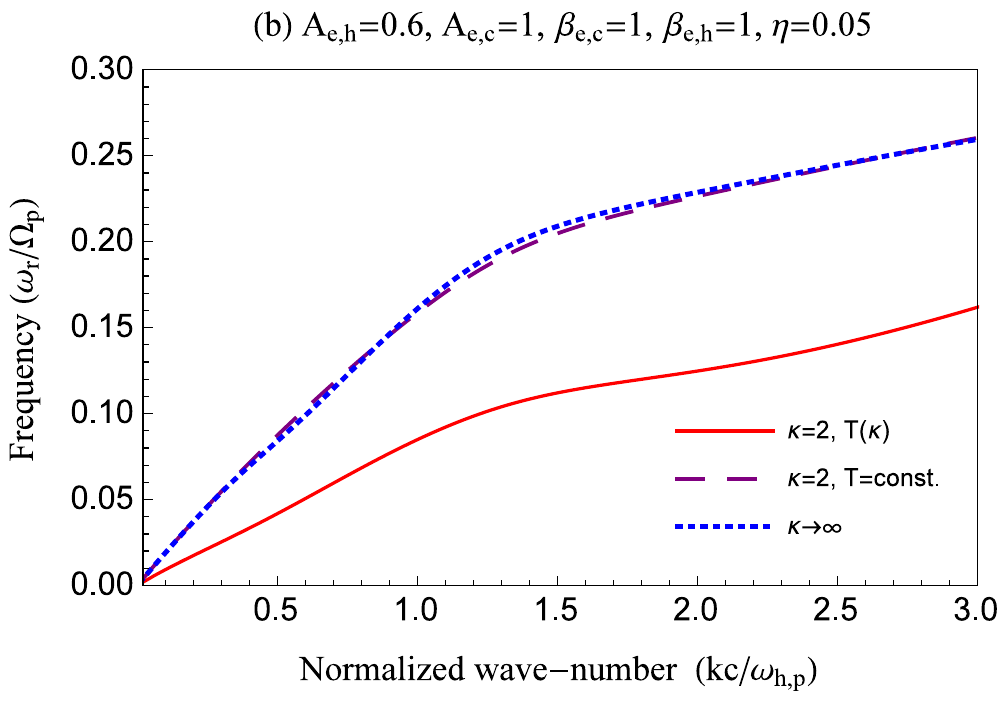}
\caption{The same as in Fig.~\ref{f6}, but for a lower plasma beta $\beta_{e,h}=1$. The plasma parameters 
for electrons are given explicitly in each panel and for protons are $A_{p,c}=A_{p,h}=1$, $\beta_{p,c}= 
\beta_{p,h}=1$.}\label{f7}
        \end{figure}

\section{Kappa electrons: destabilizing effects}\label{dest}

In this section we extend our comparative study to the effects of suprathermal 
(halo) electrons on the temperature anisotropy instabilities as they are predicted 
by the same dual Maxwellian-Kappa model when the temperatures of Kappa populations are
dependent or independent of $\kappa$. Also considered is a comparison with the Maxwellian limit 
$\kappa \to \infty$ that enables us to emphasize the effects of suprathermal electrons which 
are expected to stimulate the instability.

According to the reports on the temperature anisotropy in the solar wind \citep{Stverak-etal-2008, 
Pierrard-etal-2016} both the core and halo populations may exhibit anisotropies, and the 
halo is in general more anisotropic being hotter and less dense than the core. Therefore, 
in order to depict the effects of the suprathermal populations, here we minimize the 
influence of the Maxwellian core assuming isotropic with $T_{c,\parallel} = T_{c,\perp} = T_c$ 
(i.e., $A_c = T_{c,\perp}/T_{c,\parallel} = 1$). In the solar wind the electron halo may exhibit
both deviations from isotropy, namely, an excess of temperature in parallel direction $A_{e,h}<1$
that can drive the so-called electron firehose (EFH) instability, or an excess of temperature
in perpendicular direction $A_{e,h}>1$ that can be at the origin of the whistler instability (WI), 
also known as the electromagnetic electron-cyclotron (EMEC) instability. We have analyzed the unstable 
solutions for a significant number of cases, keeping constant the core parameters for average values 
measured in the solar wind, e.g., $A_{e,c} =1$, $\beta_{e,c} = 1$ the relative halo-core density 
$\eta=0.05$, and varying the halo parameters, i.e., $A_{e,h}$, $\beta_{e,h}$, and $\kappa$.

\subsection{Firehose instability}

We first discuss the electron firehose instability (EFHI) which is driven by a temperature anisotropy 
$A_{e,h} = T_{h,\perp} /T_{h,\parallel} < 1$ of the electron halo component. The contribution of protons 
is minimized by considering them isotropic (both the proton core and halo components), such that the dispersion 
relation for the EFHI can take the following form 
\begin{equation}
     \begin{aligned}
\tilde{k}^2=\mu \left[A_{e,h}-1+\dfrac{A_{e,h}\left(\tilde{\omega}+\mu\right)-\mu}{\tilde{k}\sqrt{\alpha^2\;\mu\;\beta_{e,h}^{\;\kappa}}}
Z_\kappa\left(\frac{\tilde{\omega}+\mu}{\tilde{k}\sqrt{\alpha^2\;\mu\;\beta_{e,h}^{\;\kappa}}}\right) \right]\\
+\dfrac{1}{\eta}\left[\dfrac{\tilde{\omega}}{\tilde{k}\sqrt{\eta\;\beta_{p,c}}}
Z_M\left( \frac{\tilde{\omega}-1}{\tilde{k}\sqrt{\eta \;\beta_{p,c}}}\right) \right]+
\dfrac{\tilde{\omega}}{\tilde{k}\sqrt{\alpha^2\;\beta_{p,h}^{\;\kappa}}}
Z_\kappa\left(\frac{\tilde{\omega}-1}{\tilde{k}\sqrt{\alpha^2\;\beta_{p,h}^{\;\kappa}}}\right)\\
+\dfrac{\mu}{\eta}\left[\dfrac{\tilde{\omega}}{\tilde{k}\sqrt{\mu\;\eta \;\beta_{e,c}}}
Z_M\left( \frac{\tilde{\omega}+\mu}{\tilde{k}\sqrt{\mu\;\eta\;\beta_{e,c}}}\right) \right] 
    \label{e18}   
     \end{aligned}     
\end{equation}
where $\tilde{k}=~kc/\omega_{h,p}$, $\tilde{\omega}=\omega /\Omega _{p}$, $\kappa=~\kappa_{e,p}$, $\alpha=(1-1.5/\kappa)^{0.5}$,
$\mu =~m_{p}/m_{e}$ is the proton-electron mass ratio, $\eta$ is the halo-core relative density,  
$A_{j,h}$, $A_{j,c}$, respectively, are, respectively, the halo and core temperature anisotropies for protons 
(subscript $j=p$) or electrons (subscript $j=e$), $\beta_{j,h}^{\;\kappa} = 8\pi nk_B T_{j,h,\parallel} / B_0^2$ 
and $\beta_{j,c} = 8\pi nk_B T_{j,c,\parallel} / B_0^2$ are the plasma
beta parameters for different plasma components, $Z_{j,M}$ is the standard dispersion function for 
(bi)-Maxwellian distributed plasmas, and $Z_{j,\kappa}$ is the modified dispersion function for (bi)-Kappa distributed plasmas  
(see Appendix A for the definitions of these functions).

%

\begin{figure}[t]
  \centering
     \includegraphics[width=0.4\textwidth]{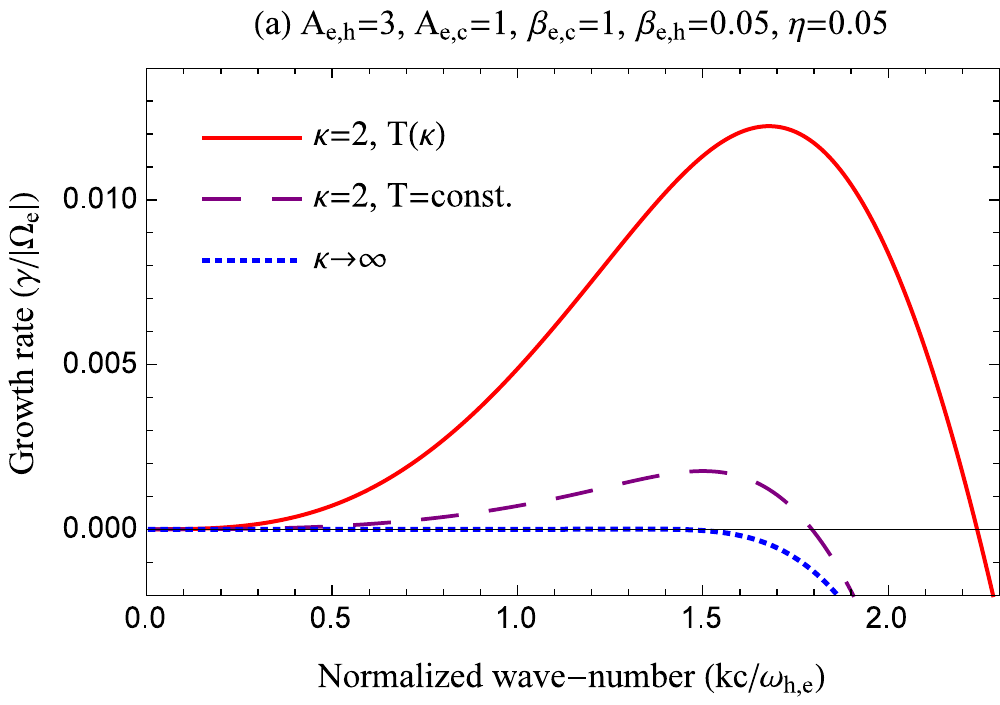}
     \includegraphics[width=0.4\textwidth]{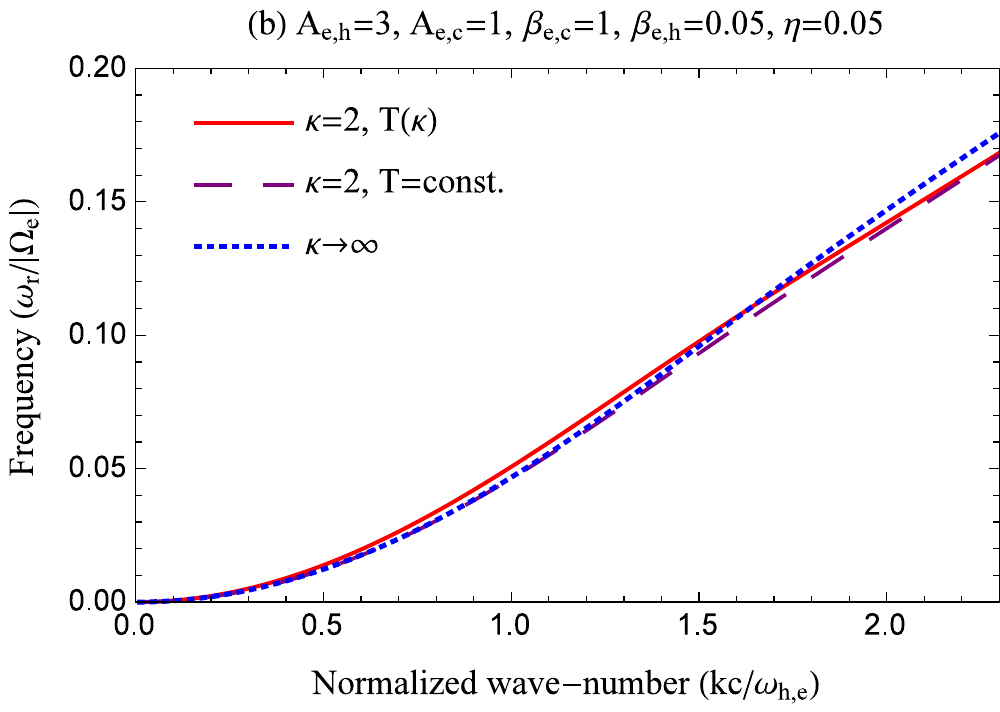}
      \caption{WI: growth rates (top) and real frequencies (bottom) derived for a dual model comprising 
			a Maxwellian core and a bi-Kappa halo ($\kappa =2$) with $\kappa-$dependent temperatures (solid lines),
			$\kappa-$independent temperatures (dashed lines), or their bi-Maxwellian limit ($\kappa \to \infty$,
			dotted-lines). The electron plasma parameters are given explicitly in each panel.}\label{f8}
  \end{figure}
\begin{figure}[t]
  \centering
     \includegraphics[width=0.4\textwidth]{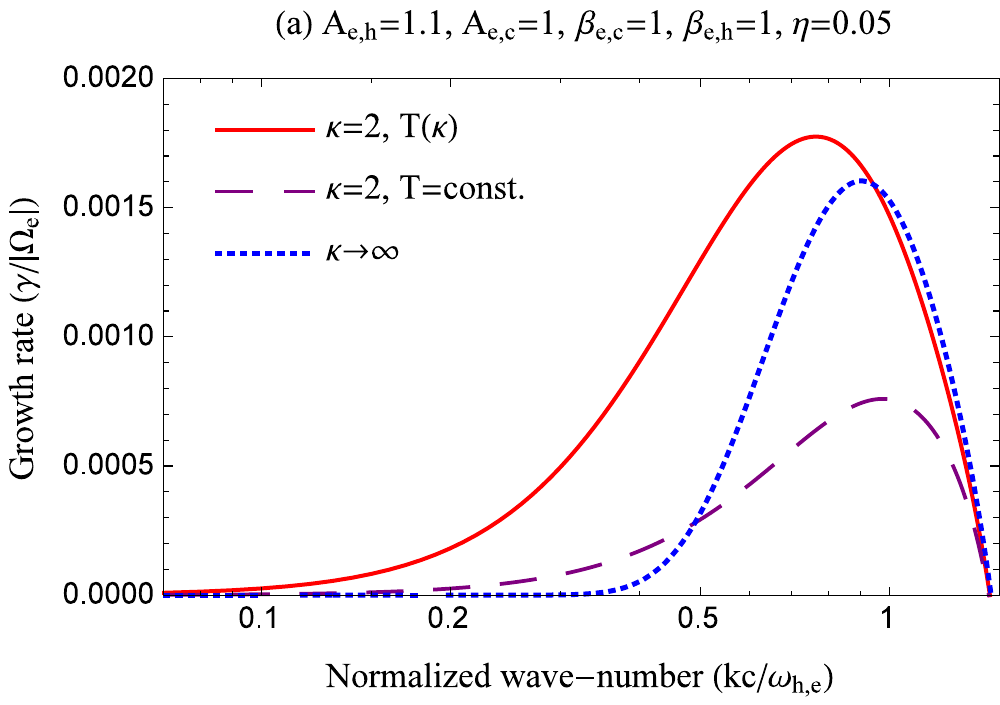}
     \includegraphics[width=0.4\textwidth]{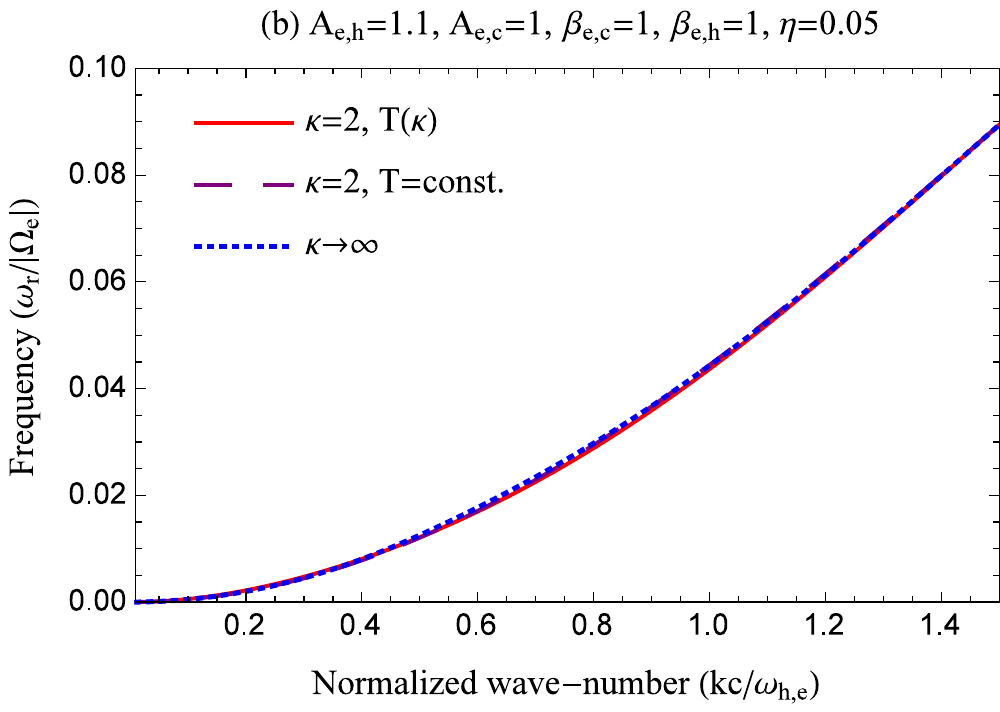}
      \caption{The same as in Fig.~\ref{f8}, but for a lower halo anisotropy $A_{e,h}=~1.1$, and a higher 
			plasma beta $\beta_{e,h}=1$.}\label{f9}
  \end{figure}

Since the plasma beta parameter is an important factor in triggering the EFH instability by the 
anisotropic electrons with $A_e <1$, the unstable solutions displayed in Figs.~\ref{f6} and \ref{f7} 
are derived for two distinct conditions, respectively, for a high halo plasma beta $\beta_{e,h}=4$, and 
for a lower $\beta_{e,h}=1$. For each of these two cases, the comparison invokes both Kappa approaches ($\kappa =2$) with 
temperatures dependent or independent of $\kappa$, as well as their Maxwellian limit ($\kappa \to \infty$). 
The growth rates (top panels) predicted by a Kappa model with $\kappa-$dependent temperatures are markedly 
enhanced in the presence of suprathermals. For instance, in Fig.~\ref{f7} only this model predicts an 
instability, while the system becomes stable within the other two approaches. A systematic stimulation 
of the EFH instability in the presence of suprathermal electrons is confirmed only by a Kappa model with 
$\kappa-$dependent temperatures. The corresponding wave-frequencies (bottom panels) also show distinct and 
important variations with the Kappa models and the power-index $\kappa$. 


\subsection{Electron cyclotron (whistler) instability }

In the opposite case an excess of perpendicular temperature $A_{e,h} = T_{h,\perp} /T_{h,\parallel} > 1$ 
may drive  the whistler instability. Since the wave frequency of the whistler modes is high ($\omega \gg 
\Omega_p$), the protons do not react and the dispersion relation that describes the whistler instability (WI) 
in a dual Maxwellian-Kappa plasma reads \citep{Lazar-etal-2015b}
\begin{equation}
     \begin{aligned}
K^2=A_{e,h}-1+\dfrac{A_{e,h}\left(W-1\right)+1}{\tilde{k}\sqrt{\alpha^2\;\beta_{e,h}^{\;\kappa}}}
Z_\kappa\left(\frac{W-1}{K\sqrt{\alpha^2\;\beta_{e,h}^{\;\kappa}}}\right)\\
+\dfrac{1}{\eta}\left[\dfrac{W}{K\sqrt{\eta \;\beta_{e,c}}}Z_M\left( \frac{W-1}{K\sqrt{\eta\;
\beta_{e,c}}}\right) \right] \label{e19}   
     \end{aligned}     
\end{equation}
where $K=kc/\omega_{h,e}$, $W=\omega /\vert\Omega _{e}\vert$, and the other quantities are the 
same as explained above for the EFHI. 

Figs.~\ref{f8} and \ref{f9} present two representative conditions for the WI, respectively,
when the halo plasma beta is low $\beta_{e,h}=0.05$ but the halo anisotropy is high $A_{e,h}=3$, 
and for a higher $\beta_{e,h}=1$ and a low anisotropy $A_{e,h}=1.1$. For even higher betas the instability 
can be triggered by very low anisotropies close to the isotropy condition $A_{e,h} \to 1$. Top panels 
display the growth-rates, and the corresponding wave-frequencies are plotted in the bottom panels.
The unstable solutions are derived for both Kappa approaches ($\kappa =2$), with temperatures dependent
or independent of $\kappa$, and for their Maxwellian limit ($\kappa \to \infty$) enabling us to emphasize 
the effects of suprathermal populations. Comparing the growth rates from different models, a systematic 
stimulation of the instability in the presence of suprathermals can be inferred only for the approach 
with $\kappa$-dependent temperatures. The other Kappa approach assuming $\kappa$-independent temperature 
leads to an irregular variation of the growth rates with $\kappa$. Unlike the growth-rates, the wave 
frequencies are not markedly influenced by the Kappa models and the variation of $\kappa$.

\section{Conclusions}

The temperature and power-index $\kappa$ are the main fitting parameters in a Kappa distribution
model widely invoked to describe the velocity distributions of plasma particles and their dynamics.
Recent studies have raised an important question on two alternative approaches, considering the
temperature of Kappa populations either dependent or independent of $\kappa$. Various scenarios can provide
justification for each of these two approaches, but observational evidences supporting any of them
are not reported yet. Here we have shown for the first time that Kappa populations observed
in the solar wind exhibit a direct interdependence between $T$ and $\kappa$. Thus, the temperature
decreases with increasing $\kappa$
\be 
T_\kappa = {\kappa \over \kappa - 3/2} T_{\kappa \to \infty}, \label{e20}
\ee
converging asymptotically to a limit that can be considered the cooler Maxwellian limit 
$T_{\kappa \to \infty} = T_M^(a)$, defined (by the components) in Eqs.~(\ref{e12})-(\ref{e13}).
This is exactly the variation predicted by the previous studies \citep{Lazar-etal-2015, Lazar-etal-2016, 
Pierrard-etal-2016}. It is important to emphasize that observational data invoked in the present paper are obtained
with a refined dual Maxwellia-Kappa model, which is more accurate than a global Kappa in reproducing 
the observed distributions and implicitly the Kappa populations. 
We have also used these two approaches with temperatures dependent or independent of $\kappa$ 
to describe and compare the effects of suprathermal electrons on two electromagnetic instabilities driven by the 
temperature anisotropy, namely, the firehose and whistler instabilities. Only the approach with $\kappa$-dependent 
temperatures may confirm the expectation predicting a systematic stimulation of the instabilities by
increasing the presence of suprathermals, while a Kappa approach with constant temperatures leads to a 
questionable variation of the growth rates with $\kappa$.



\begin{acknowledgements}
The authors acknowledge support from the Katholieke Universiteit
Leuven, the Ruhr-Universit\"at Bochum, and the Deutsche
Forschungsgemeinschaft (DFG). These results were obtained in the framework of 
the projects GOA/2015-014 (KU Leuven), G.0A23.16N (FWO-Vlaanderen) and C~90347 (ESA Prodex).
This research has been funded by the Interuniversity Attraction Poles Programme initiated 
by the Belgian Science Policy Office (IAP P7/08 CHARM). 
S.M. Shaaban would like to thank the Egyptian Ministry of Higher Education for supporting 
his research activities.
Thanks are due to \v{S}. \v{S}tver\'{a}k for providing the observational data.
\end{acknowledgements}

\section*{Appendix A: Plasma dispersion functions}

For the bi-Maxwellian distributed plasma components (e.g., the core of different species, i.e., electrons with subscript $j=e$ and protons 
with subscript $j=p$), the plasma dispersion function in Eqs.~(\ref{e18}) and (\ref{e19}) takes the standard 
form \citep{Fried-Conte-1961}
\begin{equation}
Z_{j,M}\left( \xi _{j,M}^{\pm }\right) =\frac{1}{\pi ^{1/2}}\int_{-\infty
}^{\infty }\frac{\exp \left( -x^{2}\right) }{x-\xi _{j,M}^{\pm }}dt,\
\ \Im \left( \xi _{j,M}^{\pm }\right) >0  \label{ea1}
\end{equation}
of argument $\; \xi _{j,M}^{\pm }= \left(\omega \pm \Omega_j\right)/\left(k\;\theta_{j,\parallel,}\right)$. 
For the bi-Kappa distributed components, in the dispersion relations~(\ref{e18}) and (\ref{e19}) we use the modified Kappa dispersion 
function \citep{Lazar-etal-2008} 
\begin{equation}
     \begin{aligned}
     Z_{j,\kappa}\left(\xi_{j,\kappa}^{\pm }\right)&=\frac{1}{\pi^{1/2}\kappa^{1/2}}\frac{\Gamma \left(\kappa \right)}{\Gamma \left(\kappa-1/2\right)}\\    
&\times\int_{-\infty }^{\infty}\frac{\left(1+x^{2}/\kappa\right)^{-\kappa}}{x-\xi_{j,\kappa}^{\pm }}dx,\ \Im \left(\xi _{j,\kappa}^{\pm }\right) >0,  \label{ea2}
     \end{aligned}
\end{equation}
of argument $\;\xi_{j,\kappa}^{\pm }=\left(\omega \pm \Omega _{j}\right)/\left(k\; \theta_{j,\parallel,}\right).$
In these equations $\pm$ denote the circular polarizations, right-handed (RH) and left-handed (LH), respectively, and 
$\Omega_j$ is the (non-relativistic) gyrofrequency.


\bibliographystyle{aa}

\bibliography{Kappa-obs}

\begin{thebibliography}{29}
\expandafter\ifx\csname natexlab\endcsname\relax\def\natexlab#1{#1}\fi

\bibitem[{{Christon} {et~al.}(1989){Christon}, {Williams}, {Mitchell}, {Frank},
  \& {Huang}}]{Christon-etal-1989}
{Christon}, S.~P., {Williams}, D.~J., {Mitchell}, D.~G., {Frank}, L.~A., \&
  {Huang}, C.~Y. 1989, \jgr, 94, 13409

\bibitem[{{Collier} {et~al.}(1996){Collier}, {Hamilton}, {Gloeckler},
  {Bochsler}, \& {Sheldon}}]{Collier-etal-1996}
{Collier}, M.~R., {Hamilton}, D.~C., {Gloeckler}, G., {Bochsler}, P., \&
  {Sheldon}, R.~B. 1996, \grl, 23, 1191

\bibitem[{{Feldman} {et~al.}(1975){Feldman}, {Asbridge}, {Bame}, {Montgomery},
  \& {Gary}}]{Feldman-etal-1975}
{Feldman}, W.~C., {Asbridge}, J.~R., {Bame}, S.~J., {Montgomery}, M.~D., \&
  {Gary}, S.~P. 1975, \jgr, 80, 4181

\bibitem[{Fried \& Conte(1961)}]{Fried-Conte-1961}
Fried, B.~D. \& Conte, S.~D. 1961, The plasma dispersion function: the Hilbert
  transform of the Gaussian

\bibitem[{{Gary} {et~al.}(1975){Gary}, {Feldman}, {Forslund}, \&
  {Montgomery}}]{Gary-etal-1975}
{Gary}, S.~P., {Feldman}, W.~C., {Forslund}, D.~W., \& {Montgomery}, M.~D.
  1975, \jgr, 80, 4197

\bibitem[{{Gary} {et~al.}(1994){Gary}, {Moldwin}, {Thomsen}, {Winske}, \&
  {McComas}}]{Gary-etal-1994}
{Gary}, S.~P., {Moldwin}, M.~B., {Thomsen}, M.~F., {Winske}, D., \& {McComas},
  D.~J. 1994, \jgr, 99, 23

\bibitem[{{Hellberg} {et~al.}(2005){Hellberg}, {Mace}, \&
  {Cattaert}}]{Hellberg-etal-2005}
{Hellberg}, M., {Mace}, R., \& {Cattaert}, T. 2005, \ssr, 121, 127

\bibitem[{{Lazar} {et~al.}(2016a){Lazar}, {Fichtner}, \&
  {Yoon}}]{Lazar-etal-2016}
{Lazar}, M., {Fichtner}, H., \& {Yoon}, P.~H. 2016a, \aap, 589, A39

\bibitem[{{Lazar} {et~al.}(2015a){Lazar}, {Poedts}, \&
  {Fichtner}}]{Lazar-etal-2015}
{Lazar}, M., {Poedts}, S., \& {Fichtner}, H. 2015a, \aap, 582, A124

\bibitem[{{Lazar} {et~al.}(2014){Lazar}, {Poedts}, \&
  {Schlickeiser}}]{Lazar-etal-2014}
{Lazar}, M., {Poedts}, S., \& {Schlickeiser}, R. 2014, Journal of Geophysical
  Research (Space Physics), 119, 9395

\bibitem[{{Lazar} {et~al.}(2015b){Lazar}, {Poedts}, {Schlickeiser}, \&
  {Dumitrache}}]{Lazar-etal-2015b}
{Lazar}, M., {Poedts}, S., {Schlickeiser}, R., \& {Dumitrache}, C. 2015b,
  \mnras, 446, 3022

\bibitem[{{Lazar} {et~al.}(2008){Lazar}, {Schlickeiser}, \&
  {Shukla}}]{Lazar-etal-2008}
{Lazar}, M., {Schlickeiser}, R., \& {Shukla}, P.~K. 2008, Physics of Plasmas,
  15, 042103

\bibitem[{{Lazar} {et~al.}(2016b){Lazar}, {Shaaban}, {Poedts}, \& {{\v
  S}tver{\'a}k}}]{Lazar-etal-2016b}
{Lazar}, M., {Shaaban}, S.~M., {Poedts}, S., \& {{\v S}tver{\'a}k}, {\v S}.
  2016b, \mnras

\bibitem[{{Leubner}(2004)}]{Leubner-2004}
{Leubner}, M.~P. 2004, \apj, 604, 469

\bibitem[{{Livadiotis} \& {McComas}(2013)}]{Livadiotis-McComas-2013}
{Livadiotis}, G. \& {McComas}, D.~J. 2013, \ssr, 175, 183

\bibitem[{{Maksimovic} {et~al.}(1997){Maksimovic}, {Pierrard}, \&
  {Riley}}]{Maksimovic-etal-1997}
{Maksimovic}, M., {Pierrard}, V., \& {Riley}, P. 1997, \grl, 24, 1151

\bibitem[{{Maksimovic} {et~al.}(2005){Maksimovic}, {Zouganelis}, {Chaufray},
  {Issautier}, {Scime}, {Littleton}, {Marsch}, {McComas}, {Salem}, {Lin}, \&
  {Elliott}}]{Maksimovic-etal-2005}
{Maksimovic}, M., {Zouganelis}, I., {Chaufray}, J.-Y., {et~al.} 2005, Journal
  of Geophysical Research (Space Physics), 110, A09104

\bibitem[{{Nieves-Chinchilla} \&
  {Vi{\~n}as}(2008)}]{Nieves-Chinchilla-etal-2008}
{Nieves-Chinchilla}, T. \& {Vi{\~n}as}, A.~F. 2008, Journal of Geophysical
  Research (Space Physics), 113, A02105

\bibitem[{{Pierrard} \& {Lazar}(2010)}]{Pierrard-Lazar-2010}
{Pierrard}, V. \& {Lazar}, M. 2010, Sol.\ Phys., 267, 153

\bibitem[{{Pierrard} {et~al.}(2016){Pierrard}, {Lazar}, {Poedts}, {{\v
  S}tver{\'a}k}, {Maksimovic}, \& {Tr{\'a}vn{\'{\i}}{\v
  c}ek}}]{Pierrard-etal-2016}
{Pierrard}, V., {Lazar}, M., {Poedts}, S., {et~al.} 2016, \solphys, 291, 2165

\bibitem[{{Pierrard} {et~al.}(1999){Pierrard}, {Maksimovic}, \&
  {Lemaire}}]{Pierrard-etal-1999}
{Pierrard}, V., {Maksimovic}, M., \& {Lemaire}, J. 1999, \jgr, 104, 17021

\bibitem[{{Pilipp} {et~al.}(1987){Pilipp}, {Muehlhaeuser}, {Miggenrieder},
  {Montgomery}, \& {Rosenbauer}}]{Pilipp-etal-1987}
{Pilipp}, W.~G., {Muehlhaeuser}, K.-H., {Miggenrieder}, H., {Montgomery},
  M.~D., \& {Rosenbauer}, H. 1987, \jgr, 92, 1075

\bibitem[{Shaaban {et~al.}(2016)Shaaban, Lazar, Poedts, \&
  Elhanbaly}]{Shaaban-etal-2016}
Shaaban, S.~M., Lazar, M., Poedts, S., \& Elhanbaly, A. 2016, Journal of
  Geophysical Research: Space Physics, 121, 6031, 2016JA022587

\bibitem[{{Summers} \& {Thorne}(1991)}]{Summers-Thorne-1991}
{Summers}, D. \& {Thorne}, R.~M. 1991, Physics of Fluids B, 3, 1835

\bibitem[{{{\v S}tver{\'a}k} {et~al.}(2008){{\v S}tver{\'a}k},
  {Tr{\'a}vn{\'{\i}}{\v c}ek}, {Maksimovic}, {Marsch}, {Fazakerley}, \&
  {Scime}}]{Stverak-etal-2008}
{{\v S}tver{\'a}k}, {\v S}., {Tr{\'a}vn{\'{\i}}{\v c}ek}, P., {Maksimovic}, M.,
  {et~al.} 2008, Journal of Geophysical Research (Space Physics), 113, A03103

\bibitem[{{Vasyliunas}(1968)}]{Vasyliunas-1968}
{Vasyliunas}, V.~M. 1968, \jgr, 73, 2839

\bibitem[{{Vi{\~n}as} {et~al.}(2010){Vi{\~n}as}, {Gurgiolo},
  {Nieves-Chinchilla}, {Gary}, \& {Goldstein}}]{Vinas-etal-2010}
{Vi{\~n}as}, A., {Gurgiolo}, C., {Nieves-Chinchilla}, T., {Gary}, S.~P., \&
  {Goldstein}, M.~L. 2010, Twelfth International Solar Wind Conference, 1216,
  265

\bibitem[{{Yoon}(2014)}]{Yoon-2014}
{Yoon}, P.~H. 2014, Journal of Geophysical Research (Space Physics), 119, 7074

\bibitem[{{Zimbardo} {et~al.}(2010){Zimbardo}, {Greco}, {Sorriso-Valvo},
  {Perri}, {V{\"o}r{\"o}s}, {Aburjania}, {Chargazia}, \&
  {Alexandrova}}]{Zimbardo-etal-2010}
{Zimbardo}, G., {Greco}, A., {Sorriso-Valvo}, L., {et~al.} 2010, \ssr, 156, 89

\end{thebibliography}

\end{document}